\documentstyle[preprint,aps,floats,epsfig]{revtex}

\tightenlines

\def\appendix{\par
 \setcounter{section}{0}
 \setcounter{subsection}{0}
 \def\thesection{Appendix \Alph{section}}
 \def\thesubsection{\Alph{section}.\arabic{subsection}}
 \def\theequation{\Alph{section}.\arabic{equation}}
 \setcounter{equation}{0}}

\begin{document}

\renewcommand{\thefootnote}{\fnsymbol{footnote}}

\begin{flushright}
(appears in March 01, 2001, issue of Phys.~Rev.~D)
\end{flushright}

\vspace{0.5cm} 

\centerline{{\large \bf 
Scale- and scheme--independent extension of Pad\'e
approximants;}} 
\centerline{{\large \bf
Bjorken polarized sum rule as an example}}

\vspace{1.cm}

\centerline{G.~Cveti\v c}

\centerline{{\it Asia Pacific Center for Theoretical Physics,
Seoul 130-012, Korea}}
\centerline{{\it and}}
\centerline{{\it Dept.~of Physics, Universidad T\'ecnica Federico Santa 
Mar\'{\i}a, Valpara\'{\i}so, Chile\footnote{address after August,
2000}}}
\centerline{{\it e-mail: cvetic@fis.utfsm.cl}}

\vspace{0.8cm}

\centerline{R.~K\"ogerler}

\centerline{{\it  
Dept.~of Physics, Universit\"at Bielefeld,
33501 Bielefeld, Germany}}
\centerline{{\it e-mail: koeg@physik.uni-bielefeld.de}}

\renewcommand{\thefootnote}{\arabic{footnote}}

\begin{abstract}

A renormalization--scale--invariant generalization of the diagonal 
Pad\'e approximants (dPA), developed previously, is extended so
that it becomes renormalization--scheme--invariant as well. We do
this explicitly when two terms beyond the leading order 
(NNLO,$\sim$$\alpha_s^3$) are known in the truncated perturbation 
series (TPS). At first, the scheme dependence shows up as a 
dependence on the first two scheme parameters $c_2$ and $c_3$.
Invariance under the change of the leading parameter $c_2$ is achieved 
via a variant of the principle of minimal sensitivity. The subleading 
parameter $c_3$ is fixed so that a scale-- and scheme--invariant 
Borel transform of the resummation approximant gives the correct 
location of the leading infrared renormalon pole. The leading 
higher--twist contribution, or a part of it, is thus believed 
to be contained implicitly in the resummation. We applied the 
approximant to the Bjorken polarized sum rule (BjPSR) at 
$Q^2_{\rm ph}\!=\!5$ and $3 \ {\rm GeV}^2$,
for the most recent data and the data available until 1997, 
respectively, and obtained 
${\alpha}_s^{{\overline {\rm MS}}}(M_Z^2)\!=\!0.119^
{+0.003}_{-0.006}$ and $0.113^{+0.004}_{-0.019}$, respectively.
Very similar results are obtained with the Grunberg's effective 
charge method and Stevenson's TPS principle of minimal sensitivity,
if we fix $c_3$--parameter in them by the afore-mentioned
procedure. The central values for 
${\alpha}_s^{{\overline {\rm MS}}}(M_Z^2)$
increase to $0.120$ ($0.114$) when applying dPA's,
and $0.125$ ($0.118$) when applying NNLO TPS.\\
PACS number(s): 11.10.Hi, 11.80.Fv, 12.38.Bx, 12.38.Cy

\end{abstract}

\setcounter{equation}{0}

\section{Introduction}

The problem of extracting as much information as possible,
from an available QCD or QED truncated perturbation series 
(TPS) of an observable, and including this information in a 
resummed result, was the focus of several works during the last 
twenty years. Most of these resummation methods are 
based on the available TPS only. 
Some of these latter methods eliminate the unphysical 
dependence of the TPS on the renormalization scale (RScl) 
and scheme (RSch) by fixing them in the TPS itself. 
Among these methods are the BLM fixing motivated by large--$n_f$ 
considerations \cite{BLM}, principle of minimal 
sensitivity (PMS) \cite{PMS}, effective charge method (ECH) 
\cite{ECH,KKP} (cf. Ref.~\cite{Gupta} for a related method).
Some of the more recent approaches in this direction include
approaches related with the method of ``commensurate scale
relations'' \cite{csr}, an approach using an analytic form
of the coupling parameter \cite{AQCD}, ECH--related approaches
\cite{ECHrel}, a method using expansions in the
two--loop coupling parameter \cite{Kourashev} 
expressed in terms of the Lambert function \cite{Lambert},
methods using conformal transformations either for the
Borel expansion parameter \cite{conft}
or for the coupling parameter \cite{Solovtsov}.
A basically different method consists in replacing
the TPS by Pad\'e approximants (PA's) which provide
a resummation of the TPS such that the resummed results
show weakened RScl and RSch dependence \cite{Pade}.
In particular, the diagonal Pad\'e approximants (dPA's) were
shown to be particularly well motivated since they are 
RScl--independent in the approximation of the one--loop 
evolution of the coupling ${\alpha}_s(Q^2)$ \cite{Gardi}.
An additional advantage of PA's is connected with the fact
that they surmount the purely polynomial structure of the
TPS's on which they are based, 
and thus offer a possibility of accounting for at 
least some of the nonperturbative contributions, via
a strong mechanism of quasianalytic continuation implicitly
contained in PA's.

Recently, we proposed a generalization of the
method of dPA's which achieves the exact perturbative
RScl independence of the resummed result \cite{GC1}.
While this procedure in its original form was restricted
to the cases where the number of available TPS terms beyond
the leading order (LO: $\sim$$\alpha^1$) is odd,
it was subsequently extended to the remaining cases where this
number is even \cite{CK}. This would then apply to
those QCD observables where the number of such known
terms is two (NNLO,$\sim$$\alpha_s^3$).\footnote{
When just one such term is known (NLO), our 
approximants give the same result as the ECH method.}
In \cite{CK} we also speculated on ways how to eliminate
the leading RSch--dependence from our approximants $A$, 
and proposed for the NNLO case a simple way following
the principle of minimal sensitivity (PMS). It turns out that
the way proposed there does not work properly in
practice since no minimum of the PMS equation
$\partial A/\partial c_2\!=\!0$ [cf. Eq. (40) there]
can be found. The dependence
of our approximants on the RSch--parameters 
$c_2\!\equiv\!\beta_2/\beta_0$ and 
$c_3\!\equiv\!\beta_3/\beta_0$ of the original TPS
is definitely a problem when the approximants are
applied to the low--energy
observables like the Bjorken polarized sum rule
(BjPSR) at the low momentum transfer of the virtual photon, 
e.g. $Q^2_{\rm ph}\!\approx\!3$--$5 \ {\rm GeV}^2$ \cite{GC2}.

In the present work, we address this problem.
For the NNLO TPS case, we construct in Section II
an extended version ${\cal A}$ of our approximants, 
in which the dependence on the leading RSch-parameter 
$c_2$ is successfully eliminated by application of 
a variant of PMS conditions 
$\partial {\cal A}/\partial c_2^{(j)}\!=\!0$.
This procedure can be extended in a straightforward way
to the cases where more terms are known in the
TPS, e.g. the NNNLO cases available now in QED,
but we will not discuss such cases here.
In Section III, we apply our approximant to
the BjPSR at such $Q^2_{\rm ph}$ where three quark flavors
are assumed active, e.g.
$Q^2_{\rm ph}\!\approx\!3$--$5 \ {\rm GeV}^2$.
While the approximant at this stage is an
RScl--independent and $c_2$--independent
generalization of the diagonal Pad\'e approximant
(dPA) $[2/2]$, it still contains $c_3$--dependence 
comparable to that of the ECH \cite{ECH} and TPS--PMS 
\cite{PMS} methods. Subsequently, we fix the value of $c_3$ in our,
the ECH and the TPS--PMS approximants so that PA's of 
a modified (RScl-- and RSch--independent) Borel transform 
of these approximants yield the correct location of the
leading infrared (IR) renormalon pole. 
Thus, in the approximants we implicitly use
$\beta$--functions which go beyond the last
perturbatively calculated order of the
observable (NNLO), in order to incorporate
the afore--mentioned nonperturbative information.
In Section IV we then compare the values of 
these resummation approximants with the values for the BjPSR
extracted from experiments, and obtain predictions for
$\alpha_s(M^2_Z)$. We also apply the TPS
and various PA methods of resummation
to these values of the BjPSR
and obtain higher values for
$\alpha_s(M^2_Z)$. In Section V we redo the calculations
by applying PA--type of quasianalytic continuation
for the $\beta$--functions relevant for our, ECH,
and TPS--PMS approximants.
We further address the question
of higher--twist terms. In Section VI we discuss the
obtained numerical results for $\alpha_s(M^2_Z)$, 
and Section VII contains summary and outlook.

A brief version containing a summarized description
and application of the method can be found in \cite{brief}.
In contrast to \cite{brief}, the numerical analysis
of the BjPSR in the present paper (Sections IV, V)
uses, in addition, the most recent data of the
E155 Collaboration \cite{2000E155}.

\section{Construction of $c_2$--independent approximants}

Let us consider a (QCD) observable $S$, with negligible
mass effects, which is normalized so
that its perturbative expansion takes the canonical form
\begin{equation}
S = a_0 (1 + r_1 a_0 + r_2 a_0^2 + r_3 a_0^3 + \cdots ) \ ,
\label{Scan}
\end{equation}
where $a_0\!\equiv\!\alpha_s^{(0)}/\pi$. We suppose that this
expansion is calculated within a specific RSch and using
a specific (Euclidean) RScl $Q_0$ (symbol `0' is generically
attached to the RScl and RSch parameters in the TPS) up to NNLO,
yielding as the result the TPS
\begin{equation}
S_{[2]} = a_0 ( 1 + r_1 a_0 + r_2 a_0^2) \ .
\label{S2TPS}
\end{equation}
Here, both $a_0$ and the coefficients $r_1$ and $r_2$
are RScl-- and RSch--dependent.
The coupling parameter $a\!\equiv\!{\alpha}_s/\pi$ 
evolves under the change of the
energy scale (RScl) $Q$, within the given RSch, 
according to the following renormalization group equation (RGE):
\begin{equation} 
\frac{ \partial a (\ln Q^2; c_2^{(0)}, \cdots) }
{\partial \ln (Q^2)} = 
- \beta_0 a^2 ( 1 + c_1 a + c_2^{(0)} a^2 + 
c_3^{(0)} a^3 + \cdots ) \ ,
\label{RGE}
\end{equation}
where $\beta_0$ and $c_1$ are universal quantities 
(RScl-- and RSch--invariant),\footnote{
${\beta}_0\!=\!(11\!-\!2 n_f/3)/4$, 
$c_1\!=\!(102\!-\!38 n_f/3)/(16 {\beta}_0)$,
where $n_f$ is the number of active quark flavors.}
whereas the remaining coefficients
$c_j^{(0)}$ ($j\!\geq\!2$) 
are RSch--dependent and their values can -- on the other hand --
be used to characterize the RSch.
Consequently, in (\ref{S2TPS}) the coupling
parameter $a_0$ is a function of the RScl and RSch
\begin{equation}
a_0 \equiv  a(\ln Q_0^2;c_2^{(0)},c_3^{(0)},\cdots) \ .
\label{a0def1}
\end{equation} 
The NLO and NNLO coefficients in (\ref{S2TPS}) have,
due to the RScl and RSch independence of $S$,
the following RScl and RSch dependence:
\begin{eqnarray}
r_1 & \equiv & r_1(\ln Q_0^2) = r_1(\ln {\tilde Q}^2) + 
\beta_0 \ln \left( Q_0^2/{\tilde Q}^2 \right) \ , 
\nonumber\\
r_2 & \equiv & r_2(\ln Q_0^2;c_2^{(0)}) = 
r_1^2(\ln Q_0^2) + c_1 r_1(\ln Q_0^2) - c_2^{(0)} + \rho_2 \ ,
\label{a0def2}
\end{eqnarray} 
where $\rho_2$ is RScl-- and RSch--invariant.
Although the physical quantity $S$ must be independent of
the RScl and RSch, its TPS (\ref{S2TPS}) possesses an 
unphysical dependence on RScl and RSch which
manifests itself in higher order terms
\begin{equation}
\frac{\partial S_{[2]}}{\partial \ln Q_0^2} \sim a_0^4
\sim \frac{\partial S_{[2]}}{\partial c_2^{(0)}}
\sim \frac{\partial S_{[2]}}{\partial c_3^{(0)}} \ .
\label{RSchdep}
\end{equation}
All approximants to $S$ which are based on TPS
(\ref{S2TPS}) must fulfill the Minimal Condition: 
when expanded in powers of $a_0$ to order $a_0^3$, 
they must reproduce TPS (\ref{S2TPS}).
Further, since the full $S$ is RScl- and RSch--independent,
the approximant should preferably share
this property with $S$ if it is to bring us closer
to the actual value of $S$.
The generalization of the
diagonal Pad\'e approximants developed in Ref.~\cite{GC1}
possesses full RScl independence for massless observables. 

In its original form it
is accountable only to TPS with an odd number of terms beyond
the leading order (LO: $\sim\!a^1$). Unfortunately, however, 
QCD observables have been calculated at most to the
NNLO, i.e., at best the TPS (\ref{S2TPS}) is known.
Therefore, in Ref.~\cite{CK} we have extended the method to
the cases with even numbers of terms beyond the LO, 
in particular for the TPS of the type (\ref{S2TPS}).  
Since within the present paper we are going to apply
an extended related procedure to these cases of 
$S_{[2]}$, we recapitulate briefly the main steps 
for treating a TPS of the generic form $S_{[2]}$.
The trick consisted in introducing -- in addition to $S$ -- the
auxiliary observable ${\tilde S}\equiv S*S$, which
then gets the following formal canonical form:
\begin{eqnarray}
{\tilde S} &=& (S)^2 = a_0 (0 + a_0 + R_2 a_0^2 + R_3 a_0^3 +\cdots ),
\label{Stil}
\\
{\rm where:} \quad R_2 &=& 2 r_1, \ R_3 = r_1^2 + 2 r_2, \ldots
\label{Rj}
\end{eqnarray}
${\tilde S}$ is then known formally to NNNLO
($\sim$$a^4$) and the method can thus be applied, 
yielding an approximant ${\cal A}_{S^2}^{[2/2]}$ to ${\tilde S}$.
The corresponding approximant to $S$ is
$\sqrt{A_{S^2}^{[2/2]}}$ which has the form \cite{CK}
\begin{equation}
\sqrt{ A_{\tilde S}^{[2/2]} } = 
\left \{
{\tilde \alpha}_0 \left[ 
a( \ln {\tilde Q}_1^2; c_2^{(0)}, c_3^{(0)}, \ldots )
- a( \ln {\tilde Q}_2^2; c_2^{(0)}, c_3^{(0)}, \ldots ) \right] 
\right \}^{1/2}  \left( = S_{[2]} + {\cal O}(a_0^4) \right) \ ,
\label{Aold}
\end{equation}
and it is again exactly RScl--invariant.
Here, the two scales ${\tilde Q}_j$ ($j\!=\!1,2$)
and the factor ${\tilde \alpha}_0$ are independent of the
RScl $Q_0$ and determined by the identities
\begin{eqnarray}
{ \ln ({\tilde Q}_2^2/Q_0^2) \choose 
\ln ({\tilde Q}_1^2/Q_0^2) } &=&
\frac{1}{2 {\beta}_0 } \left[ {\tilde b}_1 
\pm \sqrt{ {\tilde b}_1^2
- 4 {\tilde b}_2 } \right] \ ,
\quad
{\tilde \alpha}_0  =  
\frac{1}{ \sqrt{ {\tilde b}_1^2 - 4 {\tilde b}_2 } } \ ,
\label{param1old}
\\
{\tilde b}_1 &=& c_1 - 2 r_1 \ , \quad
{\tilde b}_2 =  - \frac{3}{2} c_1^2 + c_2^{(0)} +
c_1 r_1 + 3 r_1^2 - 2 r_2 \ .
\label{param2old}
\end{eqnarray}
If we ignore all higher than one--loop evolution effects,
i.e., if we set $c_1\!=\!0\!=\!c_2^{(0)}$ in 
(\ref{param1old})--(\ref{param2old})
and replace the two coupling parameters in (\ref{Aold})
by their one--loop evolved (from RScl $Q_0^2$ to ${\tilde Q}_j^2$)
counterparts, then the approximant (\ref{Aold}) becomes
the square root of the $[2/2]$ Pad\'e
approximant of ${\tilde S}$. This follows from general
considerations in \cite{GC1,CK}, but can also be 
verified directly in this special case.
The approximant $[2/2]_{\tilde S}^{1/2}$ preserves
the RScl--invariance only approximately (in
the one--loop RGE approximation).

Although the RScl dependence is eliminated completely
by using the approximant (\ref{Aold}), there remains
a RSch--dependence, i.e., dependence on
$c_j^{(0)}$ ($j \geq 2$). It manifests itself to a large
degree due to  
${\partial} {\tilde b}_2/{\partial} c_2^{(0)}\!\not=\!0$
(${\partial} {\tilde b}_2/{\partial} c_2^{(0)}\!=\!3$).
In Ref.~\cite{CK} we speculated that the dependence on the
leading RSch--parameter $c_2^{(0)}$ could be
eliminated by imposing the PMS condition
of local independence (cf. Eq. (40) in \cite{CK})
\begin{equation}
\frac{ d A_{\tilde S}^{[2/2]} \left(
\left\{ \ln {\tilde Q}_j^2 (c_2^{(0)}) \right\}_j ;   
c_2^{(0)},c_3^{(0)}, \ldots \right) }
{d c_2^{(0)}} {\Bigg |}_{c_3^{(0)},\ldots}   = 0 \ ,
\label{PMSold}
\end{equation}
where implicitly ``$=\!0$'' should be understood as
``$\sim$$a_0^6$'' since in general this derivative
is $\sim$$a_0^5$.
However, expansion of this expression in powers of the
coupling $a_0$ (or: any $a$) yields
\begin{equation}
\frac{ d A_{\tilde S}^{[2/2]} }{d c_2^{(0)}} 
{\Bigg |}_{c_3^{(0)},\ldots} 
= - 10 c_1 a_0^5 + {\cal O}(a_0^6) \ .
\label{PMSoldexp}
\end{equation}
This implies that the approximant (\ref{Aold}) to $S$
has no stationary (PMS) point with respect to the
RSch--parameter $c_2^{(0)}$, since the coefficient
of the leading term in the expansion of the derivative
is constant and cannot be made equal zero by a change of the RSch. 
Also actual numerical calculations for various 
observables $S$ confirm this.

Therefore, we will modify the approximant
(\ref{Aold}) so that the new one will allow
us to remove, by a PMS condition, the dependence on
the leading RSch--parameter $c_2^{(0)}$. 
This modification must, of course, be such that
the afore--mentioned Minimal Condition is satisfied
and that the RScl--invariance is preserved.
We do this in the following way. We keep the
overall functional structure of (\ref{Aold}).
However, we replace the single set of RSch--parameters 
$c_j^{(0)}$ ($j \geq 2$), which we inherited from the TPS,
by two sets of apriori arbitrary parameters $c_j^{(1)}$
and $c_j^{(2)}$ ($j \geq 2$) in the two coupling parameters,
respectively, and we also admit new values of the reference
momenta $Q_1^2$ and $Q_2^2$ 
\begin{equation}
\sqrt{ {\cal A}_{\tilde S}^{[2/2]} } =
\left\{ {\tilde \alpha} \left[ 
a( \ln Q_1^2; c_2^{(1)}, c_3^{(1)}, \ldots )
- a( \ln Q_2^2; c_2^{(2)}, c_3^{(2)}, \ldots ) \right] 
\right\}^{1/2} 
\left( = S_{[2]} + {\cal O}(a_0^4) \right) \ .
\label{Aans}
\end{equation}
The parameters $c_j^{(1)}$ and $c_j^{(2)}$ will
be appropriately fixed. They will turn out to be 
independent of the RSch--parameters $c_j^{(0)}$ and of the
RScl $Q_0^2$ of the original TPS, just like the scales $Q_1^2$ and
$Q_2^2$ and the parameter ${\tilde {\alpha}}$ will be.\footnote{ 
Parameters $c_2^{(1)}$ and $c_2^{(2)}$
will be chosen later in the Section, by following a
variant of the PMS; $c_3^{(1)}$ and $c_3^{(2)}$
will be set equal to each other and fixed in the next 
Sections.
}
We will now require $c_2^{(1)}\!\not=\!c_2^{(2)}$,
in contrast to (\ref{Aold}) which led us to the
problem (\ref{PMSoldexp}).
This requirement is not unnatural, since
the forms (\ref{Aold}) and (\ref{Aans})
have ${\tilde Q}_1^2\!\not=\!{\tilde Q}_2^2$
and $Q_1^2\!\not=\!Q_2^2$, respectively.
The two new momentum scales $Q_j$ and the parameter ${\tilde \alpha}$
in (\ref{Aans}) will be determined, in terms of
$c_k^{(j)}$'s ($k\!=\!2,3$; $j\!=\!1,2$), by expanding the 
two coupling parameters in power series of the
original coupling $a_0$ (\ref{a0def1}) and requiring that
the Minimal Condition be fulfilled, i.e., that
the power series for ${\cal A}_{S^2}^{[2/2]}$
coincides with that of
${\tilde S}$ (\ref{Stil})-(\ref{Rj}) up to
(and including) $\sim$$a_0^4$. 
For this purpose we use the expansion for the general
$a \equiv a(\ln Q^2; c_2, c_3, \ldots)$ in powers of
$a_0 \equiv a(\ln Q_0^2; c_2^{(0)}, c_3^{(0)}, \ldots)$
as obtained in Appendix A [Eqs. (\ref{aexpina0})--(\ref{xdcj})],
and apply it to as yet unspecified parameters
$Q_1^2$, $Q_2^2$ and $c_k^{(j)}$ ($j\!=\!1,2$).
The resulting expressions, when introduced into the
square of the right--hand side of (\ref{Aans}), yield an expansion
in powers of $a_0$. According to the Minimal Condition,
it should coincide with (\ref{Stil}) up to $\sim$$a_0^4$.
Comparison of the coefficients of $a_0^n$ ($n\!=\!2,3,4$)
leads to the following relations:
\begin{eqnarray}
{\rm at } \ a_0^2: \qquad 
1 & = & - {\tilde \alpha} ( x_1\!-\!x_2 ) \ , \quad
\Longrightarrow \quad 
{\tilde \alpha} = \frac{(-1)}{(x_1\!-\!x_2)} =
\frac{(-1)}{\beta_0 \ln (Q_1^2/Q_2^2)} \ .
\label{tilda1}
\\
{\rm at } \ a_0^3: \qquad
2 r_1 & = & 
- \left[ (x_1^2\!-\!x_2^2) - c_1 (x_1\!-\!x_2) + \delta c_2
\right]/(x_1\!-\!x_2) \ , 
\label{eqa03}
\\
{\rm at } \ a_0^4: \
2 r_2\!+\!r_1^2 & = &
- {\Big [} - ( x_1^3\!-\!x_2^3) + 
\frac{5}{2} c_1 (x_1^2\!-\!x_2^2) - c_2^{(0)} (x_1\!-\!x_2)
\nonumber\\
&& - 3 ( x_1 \delta c_2^{(1)}\!-\!x_2 \delta c_2^{(2)} )
+ \frac{1}{2} \delta c_3 {\Big ]}/(x_1\!-\!x_2) \ ,
\label{eqa04} 
\end{eqnarray}
where we have used the notations
\begin{eqnarray}
x_j &\equiv& \beta_0 \ln ( Q_j^2/Q_0^2 )
\ , \quad
\delta c_2^{(j)} \equiv c_2^{(j)} - c_2^{(0)} \quad (j\!=\!1,2) \ ,
\\
\label{not1}
\delta c_2 &\equiv& c_2^{(1)} - c_2^{(2)} \ , \quad
\delta c_3 \equiv c_3^{(1)} - c_3^{(2)} \ .
\label{not2}
\end{eqnarray}
Eqs.~(\ref{eqa03}) and (\ref{eqa04}) are the two equations
which determine the two scales $Q_1$ and $Q_2$ 
($\Leftrightarrow$ parameters $x_1$ and $x_2$)
as functions of $c_k^{(j)}$'s ($k\!=\!2,3$; $j\!=\!1,2$). 
In order to see that these
two scales are independent of the original
RScl ($Q_0$) and of the original RSch ($c_k^{(0)}$, $k \geq 2$),
we introduce
\begin{equation}
{\tilde x}_j \equiv 
\beta_0 \ln (Q_j^2/{\tilde {\Lambda}}^2) \quad
(j\!=\!1,2) \ ,
\label{tildx}
\end{equation}
where ${\tilde \Lambda}$ is the universal QCD scale
appearing in the Stevenson
equation (\ref{Stevenson}), so it is RScl-- and
RSch--invariant. After some algebra, we can
rewrite Eqs.~(\ref{eqa03}) and (\ref{eqa04})
as a system of equations for ${\tilde x}_j$
\begin{eqnarray}
2 \rho_1 + c_1 & = &
({\tilde x}_1\!+\!{\tilde x}_2) + 
\frac{\delta c_2}{({\tilde x}_1\!-\!{\tilde x}_2)} \ ,
\label{eqinv1}
\\
2 \rho_2 + 3 \rho_1^2 - 2 c_1 \rho_1 & = &
({\tilde x}_1^2\!+\!{\tilde x}_1 {\tilde x}_2\!+\!{\tilde x}_2^2)
- \frac{5}{2} c_1 ({\tilde x}_1\!+\!{\tilde x}_2)
+ 3 \frac{({\tilde x}_1 c_2^{(1)}\!-\!{\tilde x}_2 c_2^{(2)})}
{({\tilde x}_1\!-\!{\tilde x}_2)}
- \frac{\delta c_3}{2 ({\tilde x}_1\!-\!{\tilde x}_2)} \ ,
\label{eqinv2}
\end{eqnarray}
where $\rho_1$ and $\rho_2$ are the usual RScl-- and
RSch--invariants as defined, e.g., in \cite{PMS}\footnote{
R\c aczka \cite{Raczka} used the sum of the absolute values
of terms in ${\rho}_2$ for a formulation of criteria for acceptable
RScl's and RSch's in NNLO TPS. He concluded
that the strong RScl and RSch dependence of the NNLO TPS of the  BjPSR 
(with $n_f\!=\!3$) presents a serious practical problem.}
[cf.~also (\ref{a0def2})]
\begin{eqnarray}
\rho_1 & = & \beta_0 \ln (Q_0^2/{\tilde \Lambda}^2) - r_1 \ ,
\label{rho1}
\\
\rho_2 & = & r_2 - r_1^2 - c_1 r_1 + c_2^{(0)} \ .
\label{rho2}
\end{eqnarray}
Therefore, Eqs.~(\ref{eqinv1})--(\ref{eqinv2})
show the following: If $c_2^{(1)}$ and $c_2^{(2)}$
and $\delta c_3\!\equiv\!c_3^{(1)}\!-\!c_3^{(2)}$
are chosen and fixed,
then the solutions ${\tilde x}_j$ and thus the scales $Q_j$
($j\!=\!1,2$) are independent of the RScl ($Q_0$)
and of the RSch ($c_2^{(0)}, c_3^{(0)}, \ldots$).
Thus, we have
\begin{equation}
Q_j^2 = Q_j^2(c_2^{(1)}, c_2^{(2)}; \delta c_3) \quad 
(j\!=\!1,2) \ , \quad 
{\tilde \alpha}=\frac{(-1)}{\beta_0 \ln(Q_1^2/Q_2^2)} =
{\tilde \alpha}(c_2^{(1)}, c_2^{(2)}; \delta c_3) \ .
\label{invres}
\end{equation}
Therefore, our approximant (\ref{Aans}) will be regarded
from now on as a function of only $c_k^{(j)}$ parameters
($k \geq 2$; $j\!=\!1,2$): 
${\cal A}_{S^2}^{[2/2]}(c_2^{(1)}, c_2^{(2)}; 
c_3^{(1)}, c_3^{(2)}; \ldots)$.
For actually solving the equations for the scales
$Q_1$ and $Q_2$, it is more convenient
to use Eqs.~(\ref{eqa03})--(\ref{eqa04}). For the
subsequent use, we rewrite them in the following form:
\begin{eqnarray}
y_{-}^4 -  y_{-}^2 z_0^2(c_2^{(s)}) +
y_{-} \frac{1}{4} ( 5 c_1 \delta c_2 - \delta c_3) 
- \frac{3}{16} (\delta c_2)^2 &=&  0 \ ,
\label{eq1}
\\
- r_1 + \frac{1}{2} c_1 - \frac{1}{4} \frac{\delta c_2}{y_{-}}
& = & y_{+} \ ,
\label{eq1b}
\end{eqnarray}
where we use the notations
\begin{eqnarray}
y_{\pm} &\equiv& \frac{1}{2} \beta_0 \left[ \ln \frac{Q_1^2}{Q_0^2}
\pm \ln \frac{Q_2^2}{Q_0^2} \right] \ ,
\label{notf1}
\\
\delta c_k &\equiv& c_k^{(1)} - c_k^{(2)} \ , \quad
c_k^{(s)} \equiv 
\frac{1}{2}(c_k^{(1)} + c_k^{(2)}) \quad (k\!=\!2,3) \ ,
\label{notf2}
\\
z_0^2 &\equiv& \left( 2 \rho_2\!+\!\frac{7}{4} c_1^2 \right) 
- 3 c_2^{(s)} \equiv z_0^2(c_2^{(s)}) \ ,
\label{z0}
\end{eqnarray}
where $\rho_2$ is given by (\ref{rho2}).
Incidentally, it can be explicitly checked that
in the special case of 
$c_2^{(1)}\!=\!c_2^{(2)}\!=\!c_2^{(0)}$ and
$c_3^{(1)}\!=\!c_3^{(2)}\!=\!c_3^{(0)}$
Eqs. (\ref{eq1})--(\ref{z0}) and (\ref{tilda1})
recover the old approximant (\ref{Aold})--(\ref{param2old})
of Ref.~\cite{CK}.

The next question is how to fix parameters
$c_2^{(j)}$ and $c_3^{(j)}$ ($j\!=1,2$).
Above all, we have to fix the leading parameters 
$c_2^{(j)}$'s since otherwise their arbitrariness would
reflect the fact that the leading RSch--dependence 
(i.e., the dependence on $c_2^{(0)}$)
has not been eliminated from the approximant.
We do this by requiring the local independence of
the approximant with respect to variation 
of $c_2^{(1)}$ and of $c_2^{(2)}$ separately.
This condition is a variant
of the principle of minimal sensitivity (PMS),
or a PMS--type ansatz
\begin{equation}
\frac{ \partial {\cal A}_{\tilde S}^{[2/2]} }
{\partial c_2^{(1)} } {\Bigg |}_{c_2^{(2)}}
= 0 =
\frac{ \partial {\cal A}_{\tilde S}^{[2/2]} }
{\partial c_2^{(2)} } {\Bigg |}_{c_2^{(1)}}
\quad \Longleftrightarrow \quad
 \frac{ \partial {\cal A}_{\tilde S}^{[2/2]} }
{\partial c_2^{(s)} } {\Bigg |}_{\delta c_2}
= 0 =
\frac{ \partial {\cal A}_{\tilde S}^{[2/2]} }
{\partial (\delta c_2) } {\Bigg |}_{c_2^{(s)}}
\label{PMSeqs}
\end{equation}
Here, ``$=\!0$'' should be understood as ``$\sim$$a_0^6$''
since in general these derivatives are $\sim$$a_0^5$.
These two equations then give us solutions
for the leading parameters $c_2^{(1)}$ and $c_2^{(2)}$,
once the values of the subleading parameters
$c_3^{(s)}\!\equiv\!(c_3^{(1)}\!+\!c_3^{(2)})/2$
and $\delta c_3\!\equiv\!c_3^{(1)}\!-\!c_3^{(2)}$
have been chosen.\footnote{
Also a value of $\delta c_4\!\equiv\!c_4^{(1)}\!-\!c_4^{(2)}$
has to be chosen -- see later.}
However, using Eq.~(\ref{derc3})
and the fact that $Q_j^2$ are independent of
$c_3^{(s)}$ [cf.~(\ref{invres})], we can show 
the following dependence of the approximant on
$c_3^{(s)}$ (at constant $\delta c_3$):
\begin{equation}
d \ln \left( \sqrt{ {\cal A}_{\tilde S}^{[2/2]}} \right) 
= d (c_3^{(s)}) \frac{1}{4} 
( a_1^3\!+\!a_1^2 a_2\!+\!a_1 a_2^2\!+\!a_2^3 ) 
+ {\cal O}(a_j^4)
\stackrel{<}{\approx} d (c_3^{(s)}) |a_1|^3 \ ,
\label{dAdc3}
\end{equation}
where $a_j\!\equiv\!a(\ln Q_j^2; c_2^{(j)}, c_3^{(j)},\ldots)$
($j\!=\!1,2$) and we took the index convention
$|a_1|\!\geq\!|a_2|$.
This means that the dependence on $c_3^{(s)}$
cannot be eliminated in the considered case,
not even by a PMS variant. In this respect, the situation 
is analogous to the usual TPS--PMS \cite{PMS} and the 
ECH \cite{ECH} methods. These two methods (cf. Appendix C), 
while fixing RScl ($Q_0 \mapsto Q_{\rm ECH}\!=\!Q_{\rm PMS}$)
and $c_2$ RSch--parameter 
($c_2^{(0)} \mapsto c_2^{\rm PMS}$ or $c_2^{\rm ECH}$)
in the original TPS (\ref{S2TPS}), leave the
value of the subleading parameter $c_3$ there unspecified,
with the residual $c_3$--dependence of the
(TPS--)approximant
\begin{equation}
d \ln \left( S_{[2]}^{(\rm X)} \right) 
\approx d (c_3) a_{X}^3/2  \ ,
\label{dAdc3TPS}
\end{equation}
where label `X' stands either for `ECH' of `TPS--PMS'.
Comparing (\ref{dAdc3}) and (\ref{dAdc3TPS}),
we see that the $c_3^{(s)}$--dependence of our
approximant could be up to twice as strong as that
of the TPS--PMS and ECH methods. 

Hence, varying $c_3^{(1)}$ and $c_3^{(2)}$
parameters in our approximant at this point
would apparently not lead to any new insight.
For the sake of simplicity, we choose from
now on these two subleading parameters to be equal
to each other
\begin{equation}
c_3^{(1)} = c_3^{(2)} \equiv c_3 \qquad  (\delta c_3 = 0) \ ,
\label{dc30}
\end{equation}
but we will adjust the
common parameter $c_3$ later to a physically
motivated value.

With the chosen restriction (\ref{dc30}), the problem 
of finding our approximant (\ref{Aans}) to the TPS (\ref{S2TPS}) 
basically reduces to the problem of solving the system of 
three coupled equations (\ref{eq1}) and (\ref{PMSeqs}) 
for the three unknowns $y_{-}$ [$=\!\beta_0 \ln (Q_1/Q_2)$] 
and $\delta c_2$ and $c_2^{(s)}$ 
($\Leftrightarrow c_2^{(1)}$ and $c_2^{(2)}$). 
For completeness, the PMS--like equations 
(\ref{PMSeqs}), when $\delta c_3\!=\!0\!=\!\delta c_4$,
are written explicitly in Appendix B, to the relevant order
$\sim$$a_0^5$ at which we solve them -- 
Eqs. (\ref{PMSeq1})--(\ref{PMSeq2}). 
{}From there and from (\ref{eq1}) we explicitly 
see that these three equations contain only the 
three unknowns ($y_{-}$, $c_2^{(s)}$ and $\delta c_2$)
and the (known) RScl-- and RSch--invariants
$\rho_2$ (\ref{rho2}) and $c_1\!=\!\beta_1/\beta_0$.
Interestingly enough, these three equations do not
depend on $c_3$ ($=\!c_3^{(1)}\!=\!c_3^{(2)}$).
In addition, they do not depend on any other
higher order parameters $c_k^{(j)}$ 
($k \geq 4$; $j\!=\!1,2$) appearing in 
$a_j\!\equiv\!a(\ln Q_j^2; c_2^{(j)},c_3,c_4^{(j)},\ldots)$,
except on $\delta c_4\!\equiv\!c_4^{(1)}\!-\!c_4^{(2)}$
which was taken to be zero in Eqs. (\ref{PMSeq1})--(\ref{PMSeq2}).
Hence, $Q_j$ and $c_2^{(j)}$ ($j\!=\!1,2$) will
be functions of $\rho_2$ and $c_1$ only, thus
explicitly RScl-- and RSch--invariant.
For simplicity, we want the solutions $Q_j^2$ and $c_2^{(j)}$
($j\!=\!1,2$) to be independent of {\em any\/} higher
order parameter $c_k^{(j)}$ ($k \geq 3$) that possibly appears
in our approximant, therefore we choose from now on also 
$\delta c_4$($\equiv\!c_4^{(1)}\!-\!c_4^{(2)}$)$=\!0$.
The solution of the mentioned three coupled
equations in any specific case can be found numerically, 
e.g. by using Mathematica or some other comparable software
for numerical iteration. Certainly we have to
ensure that the program scans through a sufficiently
wide range of the initial trial values $y_{-}^{(\rm in.)}$,
$(c_2^{(s)})^{(\rm in.)}$ and $(\delta c_2)^{(\rm in.)}$ 
for iterations, in order not to miss any solution.
The solutions which result in either $|{\tilde \alpha}|\!\gg\!1$
or $|{\tilde \alpha}|\!\ll\!1$ should be
discarded since they signal numerical instabilities
of the approximant [$|{\tilde \alpha}|\!\gg\!1 \Rightarrow 
Q_1^2\!\approx\!Q_2^2$ -- cf.~(\ref{tilda1})] or are in addition
physically unacceptable ($|{\tilde \alpha}|\!\ll\!1 \Rightarrow 
Q_1^2\!\ll\!Q_2^2 \ {\rm or} \ Q_2^2\!\ll\!Q_1^2 $).
We have apparently two possibilities:
\begin{itemize}
\item
$y_{-}$, $c_2^{(s)}$ and $\delta c_2$ are all real numbers
(and thus the intial trial values as well);
\item
$c_2^{(s)}$ and its initial values are real;
$y_{-}$ and $\delta c_2$ and their initial values are
imaginary numbers ($c_2^{(1)}$ and $c_2^{(2)}$ are
complex conjugate to each other, as are $Q_1^2$ and $Q_2^2$).
\end{itemize}
In both cases, the approximant itself turns out to be real,
as long as $c_3$ is real.

If we encounter several solutions which
give different values for the approximant, we should
choose, again within the PMS--logic, among them the
solution with the smallest curvature with respect
to $c_2^{(1)}$ and $c_2^{(2)}$. For such cases, we 
define two almost equivalent expressions for such curvature
in Appendix B -- cf. Eqs. (\ref{curv12})--(\ref{curv}).

\section{Bjorken polarized sum rule (BPSR): $c_3$-fixing}

We will now apply the described method to
the case of the Bjorken polarized sum rule (BjPSR) \cite{Bjorken}.
It is the isotriplet combination of the 
first moments over $x_{\rm Bj}$ of proton and
neutron polarized structure functions
\begin{equation}
\int_0^1 d x_{\rm Bj} \left[ g_1^{(p)} (x_{\rm Bj}; Q^2_{\rm ph})
- g_1^{(n)} (x_{\rm Bj}; Q^2_{\rm ph}) \right] =
 \frac{1}{6} |g_A| \left[ 1 - S(Q^2_{\rm ph}) \right] \ ,
\label{BjPSR1}
\end{equation}
where $p^2\!=\!-Q^2_{\rm ph}$$<0$ is the momentum transfer
carried by the virtual photon. The quantity $S(Q^2_{\rm ph})$ has the
canonical form (\ref{Scan}). It has been calculated to
the NNLO \cite{GLZN,LV}, 
in the ${\overline {\rm MS}}$ RSch and with the RScl 
$Q_0^2=Q^2_{\rm ph}$. The pertaining values of
$r_1$ and $r_2$, for those $Q^2_{\rm ph}$ where three
quark flavors are assumed active ($n_f\!=\!3$), e.g. at
$Q^2_{\rm ph}\!=\!3$ or $5 \ {\rm GeV}^2$, are
$r_1 = 3.5833$ \cite{GLZN} and $r_2= 20.2153$ \cite{LV},
so that
\begin{eqnarray}
S_{[2]}(Q^2_{\rm ph}; Q^2_0 = Q^2_{\rm ph}; 
c_2^{{\overline {\rm MS}}}, c_3^{{\overline {\rm MS}}})
&=& a_0 ( 1 + 3.5833 a_0 + 20.2153 a_0^2) \ ,
\label{TPSBj}
\\
{\rm with:} \quad a_0 = 
a(\ln Q_0^2; c_2^{{\overline {\rm MS}}}, 
c_3^{{\overline {\rm MS}}}, \ldots) \ , \quad 
n_f&=&3 \ , \ c_2^{{\overline {\rm MS}}} = 4.471, \ 
c_3^{{\overline {\rm MS}}} = 20.99 \ .
\label{TPSBjnot}
\end{eqnarray}
The constant $|g_A|$ appearing in (\ref{BjPSR1})
is known from $\beta$--decay measurements \cite{PDG}
(it is denoted there as $|g_A/g_V|$)
\begin{equation}
|g_A| = 1.2670 \pm 0.0035 \ .
\label{gA}
\end{equation}
Solving the coupled system of (\ref{eq1}) and
(\ref{PMSeq1})--(\ref{PMSeq2}) for the three
unknowns $y_{-}$, $c_2^{(s)}$ and $\delta c_2$,
as discussed in the previous Section, results in this
case in one physical solution only\footnote{
Formally, we get two solutions, but they give the same
approximant, since the second solution is obtained from
the first by $Q_1 \leftrightarrow Q_2$ and 
$c_2^{(1)}  \leftrightarrow c_2^{(2)}$.
Further, if ignoring in PMS conditions 
(\ref{PMSeq1})--(\ref{PMSeq2}) the denominators, one arrives
at two additional solutions, both having 
$c_2^{(s)}=(6 {\rho}_2\!-\!7 c_1^2/4)/7$; however,
one can check that also the denominators are then zero
and the derivative (\ref{PMSeq1}) reduces to
$2(2 \delta c_2\!-\!15 c_1 y_{-}){\bar a}_0^5/(3 y_{-})$
which turns out to be finite and nozero.}
\begin{eqnarray}
y_{-} {\bigg (} &\equiv& \frac{1}{2} \beta_0 \ln \frac{Q_1^2}{Q_2^2}
{\bigg )} =  - 1.514  \quad (\Rightarrow {\tilde \alpha} = 0.3301) \ ,
\label{ymres}
\\
c_2^{(s)} & = & 3.301 \ , \ \delta c_2 = - 3.672 \quad \Rightarrow
\quad c_2^{(1)} = 1.465 \ , \ c_2^{(2)} = 5.137 \ .  
\label{c2res}
\end{eqnarray}
Parameter $y_{+}$, defined in (\ref{notf1}), is then obtained
from (\ref{eq1b}). The resulting scales $Q_1$, $Q_2$ are then
$0.767$ GeV, $1.504$ GeV ($Q^2_{\rm ph}\!=\!5 \ {\rm GeV}^2$)
and $0.594$ GeV, $1.165$ GeV ($Q^2_{\rm ph}\!=\!3 \ {\rm GeV}^2$). 
We stress that these results are independent of the value of $c_3$
(\ref{dc30}) and of $c_4$ and other $c_k^{(j)}$ 
($k \geq 5$; $j\!=\!1,2$) in the approximant $\sqrt{{\cal A}_{S^2}}$
(\ref{Aans}), and are independent of the choice of RScl
$Q_0$ and RSch ($c_k^{(0)}$, $k \geq 2$)
in the original TPS $S_{[2]}$. In TPS (\ref{TPSBj}),
the choice was $Q_0\!=\!Q_{\rm ph}$
and $c_2^{(0)}\!=\!c_2^{{\overline {\rm MS}}}$ ($=\!4.471$).
Knowing $Q_j$ and $c_2^{(j)}$ ($j\!=\!1,2$), 
for the actual evaluation of approximant (\ref{Aans}) 
we need to assume a certain value for $a_0$ (\ref{TPSBjnot})
(at RScl $Q_0$). The value of
${\tilde \alpha}$ is obtained from
(\ref{tilda1}) (${\tilde \alpha}\!=\!0.3303$);
the value of the coupling parameter 
$a_j\!\equiv\!a(\ln Q_j^2; c_2^{(j)}, c_3, c_4, c_5^{(j)}, \ldots)$ 
($j\!=\!1,2$) can be obtained, for example, by solving the subtracted
Stevenson equation (\ref{Stsubtr})
\begin{eqnarray}
\lefteqn{
\beta_0 \ln \frac{Q_j^2}{Q_0^2} =
\frac{1}{a_j} + c_1 \ln \left( \frac{ c_1 a_j}{1\!+\!c_1 a_j} \right)
+ \int_0^{a_j}\!dx 
\frac{ (c_2^{(j)}\!+\!c_3 x) }{( 1\!+\!c_1 x)
(1\!+\!c_1 x\!+\!c_2^{(j)} x^2\!+\!c_3 x^3)}
}
\nonumber \\
&& - \frac{1}{a_0} 
- c_1 \ln \left( \frac{ c_1 a_0}{1\!+\!c_1 a_0} \right)
- \int_0^{a_0}\!dx
\frac{ (c_2^{{\overline {\rm MS}}}\!+\!c_3^{{\overline {\rm MS}}} x)}
{( 1\!+\!c_1 x)
(1\!+\!c_1 x\!+\!c_2^{{\overline {\rm MS}}} 
x^2\!+\!c_3^{{\overline {\rm MS}}} x^3)} 
\quad (j\!=\!1,2) \ .
\label{Stforaj}
\end{eqnarray}
In (\ref{Stforaj}) we ignored terms $\propto$$c_4^{(j)}$ and
higher since they are not known 
($c_4^{{\overline {\rm MS}}}$ is not known, either).
Stated otherwise, we set here and in the rest of this
Section: $c_k^{(1)}\!=\!c_k^{(2)}\!=\!c_k^{{\overline {\rm MS}}}\!=\!0$
for $k \geq 4$, i.e. the $\beta$--functions pertaining to
the approximant are taken in the TPS form to the four--loop order.
Hence, the only free parameter in the approximant 
$\sqrt{{\cal A}_{S^2}}$ (\ref{Aans}) is now
$c_3$ [cf. condition (\ref{dc30})], all the other
nonzero parameters ($Q_j^2$, $c_2^{(j)}$, ${\tilde \alpha}$)
have been determined and are $c_3$-- and RScl-- and RSch--independent.
Further, any effects due to the mass thresholds
($n_f \geq 4$) are ignored in (\ref{Stforaj}). 
These effects are suppressed because the difference of the
two integrals in (\ref{Stforaj}) tends to cancel them.
Note that the scales appearing in (\ref{Stforaj})
($Q_1\!\approx\!0.6$--$0.8$ GeV, $Q_2\!\approx\!1.2$--$1.5$ GeV,
$Q_{\rm ph}\!=\!Q_0\!\approx\!1.7$--$2.2$ GeV) are all regarded to be below
the threshold $(n_f\!=\!3) \mapsto (n_f\!=\!4)$,
i.e. all the active quark 
flavors are (almost) massless.\footnote{
In the whole paper, we ignore any quark mass effects, 
except later in the evolution
${\alpha}_s^{{\overline {\rm MS}}}(Q_{\rm ph}^2) \mapsto
{\alpha}_s^{{\overline {\rm MS}}}(M_Z^2)$ where the
quark mass thresholds are significant and accounted for.}

The main question appearing at this point is: Which value
of $c_3$ ($=\!c_3^{(1)}\!=\!c_3^{(2)}$) should we choose
in our approximant? The two most obvious possibilities are
$c_3\!=\!0$ or $c_3\!=\!c_3^{{\overline {\rm MS}}}$ $(=\!20.99)$.
The decision is far from being numerically irrelevant. If choosing
for $a_0\!\equiv\!a(\ln Q_0^2; c_2^{{\overline {\rm MS}}}, 
c_3^{{\overline {\rm MS}}})$
at $Q_{\rm ph}^2\!=\!Q_0^2\!=\!3$ GeV a typical value, 
e.g., $a_0\!=\!0.09$
[$\Rightarrow$
$\alpha_s^{{\overline {\rm MS}}}(3 {\rm GeV}^2)\!\approx\!0.283$,
$\alpha_s^{{\overline {\rm MS}}}(M_Z^2)\!\approx\!0.113$],
we obtain the following resummed values for the BjPSR $S$
\begin{equation}
\sqrt{ {\cal A}_{\tilde S}^{[2/2]}(c_3\!=\!0) } =  0.1523 \ ,
\qquad
\sqrt{ {\cal A}_{\tilde S}^{[2/2]}
(c_3\!=\!c_3^{{\overline {\rm MS}}}) }= 0.1632 \ .
\label{resumc3}
\end{equation}
The latter is $7.16 \%$ higher than the former.
The corresponding resummed values of the ECH \cite{ECH}
and TPS--PMS \cite{PMS} are
\begin{eqnarray}
{\cal A}_S^{\rm ECH}(c_3\!=\!0) & = & 0.1535 \ , \qquad
{\cal A}_S^{\rm ECH}(c_3\!=\!c_3^{{\overline {\rm MS}}}) 
=  0.1593 \ ;
\label{ECHc3}
\\
{\cal A}_S^{\rm PMS}(c_3\!=\!0) & = &   0.1528 \ , \qquad 
{\cal A}_S^{\rm PMS}(c_3\!=\!c_3^{{\overline {\rm MS}}})   
=   0.1588 \ .
\label{PMSc3}
\end{eqnarray}
The latter values (for $c_3\!=\!c_3^{{\overline {\rm MS}}}$) are
$3.79 \%$ (ECH) and $3.96 \%$ (TPS--PMS) higher than
the former (for $c_3\!=\!0$). Thus, the sensitivity
of our approximant to the variation of $c_3$ 
is in the considered case almost twice as large as for
the ECH and TPS--PMS methods, as anticipated in 
(\ref{dAdc3})--(\ref{dAdc3TPS}) in the previous Section. 
The true value of $c_3$ in ${\cal A}_S^{\rm ECH}$
should be equal to ${\rho}_3$, i.e. the third
RScl-- and RSch--invariant of the BjPSR, but this value
is not exactly known because the ${\rm N}^3 {\rm LO}$
coefficient $r_3$ in the perturbative expansion of the
BjPSR is not known yet.
The stronger $c_3$--sensitivity should
not be regarded as a negative feature of our approximant,
but rather within the following context:

Our approximant contains two (RScl--invariant) energy scales
$Q_1$, $Q_2$. Since the considered observable
is close to the nonperturbative sector ($Q_{\rm ph} < 2.5$ GeV),
the relevant scales $Q_j$ ($\sim$$Q_{\rm ph}$)
are low: $Q_1\!\approx\!0.6$--$0.8$ GeV and 
$Q_2\!\approx\!1.2$--$1.5$ GeV.
Thus the relevant coupling parameters 
$a_j\!\equiv\!a(\ln Q_j^2; c_2^{(j)}, c_3)$ are large:
$a_1\!\approx\!0.19$ and $a_2\!\approx\!0.11$
(when $c_3$ is set equal 
$c_3^{{\overline {\rm MS}}}$ and $a_0\!=\!0.09$,
$Q_0^2\!=\!Q_{\rm ph}^2\!=\!3 \ {\rm GeV}^2$).
Therefore, the contribution
of the $c_3$--term on the right--hand side of the integrated
RGE (\ref{Stforaj})
[$\Leftrightarrow$ differential RGE (\ref{RGE})]
at such energy scales is not negligible. This feature,
to a somewhat lesser degree, can also be seen in
the ECH and TPS--PMS approaches, where $Q_{\rm ECH}$
($=\!Q_{\rm PMS}$) $\approx\!0.8$ GeV and
$a_{\rm ECH}\!\equiv\!a(\ln Q^2_{\rm ECH}; c_2^{\rm ECH}, c_3)
\approx 0.16$ (when $c_3$ is set equal to 
$c_3^{{\overline {\rm MS}}}$, and $a_0\!=\!0.09$,
$Q_0^2\!=\!Q_{\rm ph}^2\!=\!3 \ {\rm GeV}^2$).
The significant $c_3$--dependence of all
these approximants, at fixed $a_0$, reflects the fact that 
the coupling parameters $a(Q_j)$ appearing in the
approximants are not small and that consequently the
considered observable is in the low--energy regime.
The values of Pad\'e approximants (PA's), when applied
to NNLO TPS of an observable (e.g., BjPSR),
are also $c_3$--dependent. However, the latter
$c_3$--dependence, in contrast to that in the afore--mentioned
approximants, is not playing a highlighted role, 
since the PA's depend in addition on the leading RSch--parameter
$c_2$ ($\Leftrightarrow c_2^{(0)}$)
and even on the RScl $Q_0^2$.

The above considerations, however, do not address
the important problem presented by (\ref{resumc3}):
Which value of parameter $c_3$ should we
use in our approximant?

We note that $c_3$ characterizes the ${\rm N}^3 {\rm LO}$
term in the corresponding $\beta$--function (\ref{RGE}),
and the information on its value in a considered approximant
cannot be obtained from the NNLO TPS on which the
approximant is based. To determine the optimal value
of $c_3$ in an approximant (our, ECH, or TPS--PMS),
an important known piece of (nonperturbative) information
beyond the NNLO TPS should be incorporated into the
approximant. There are at least two natural candidates
for this: the location of the leading infrared (${\rm IR}_1$) 
and ultraviolet (${\rm UV}_1$) renormalon poles, i.e., 
the positive and negative poles of the Borel transform $B_S(z)$ 
of the observable closest to the origin
(-- for a review on renormalons, see \cite{Beneke:1999ui}). 
In the case of the BjPSR, these two locations
are known from large--${\beta}_0$ (large--$n_f$)
considerations \cite{BK,TM}: $z_{\rm pole}\!=\!1/{\beta}_0$
(${\rm IR}_1$), $z_{\rm pole}\!=\!-1/{\beta}_0$ (${\rm UV}_1$).

Which of the two leading renormalons is numerically more 
important in the BjPSR case?
In the simple Borel transform
of the BjPSR, with ${\overline {\rm MS}}$ RSch and
RScl $Q_0\!=\!Q_{\rm ph}$ ($n_f\!=\!3$), 
the ratio of the residues of the ${\rm IR}_1$ and 
${\rm UV}_1$ poles in the large--${\beta}_0$ 
approximation is $2 \exp(10/3)\!\approx\!56\!\gg\!1$
\cite{BK,Beneke:1999ui}. 
This would suggest strong numerical dominance 
of the ${\rm IR}_1$ over ${\rm UV}_1$.
However, when using there the V--scheme \cite{BLM},
i.e. ${\overline {\rm MS}}$ with RScl
$Q_0\!=\! Q_{\rm ph} \exp(-5/6)$ ($\approx\!Q_{\rm ECH}$),
this ratio goes down to $2$. This would suggest
that the ${\rm UV}_1$ (vis--\`a--vis ${\rm IR}_1$) 
is not entirely negligible. The authors of Ref.~\cite{TM}
used the 't Hooft RSch and varied the RScl
in such an approach (large--${\beta}_0$, simple Borel transform,
principal value prescription),
and their Fig.~2 for the BjPSR at 
$Q^2_{\rm ph}\!=\!2.5 \ {\rm GeV}^2$ suggests that
IR renormalon contributions to $S(Q^2_{\rm ph})$ are 3--4 times 
larger than those of the UV renormalons.
The relative strength of the UV vs IR renormalon
contributions, in the RScl-- and RSch--noninvariant approach
with simple Borel transform, appears to depend in practice 
on the choice of the RScl and RSch. 
Incidentally, a consideration of the status of the
renormalon contributions and of their scheme--dependence
was made in Ref.~\cite{KP}.
The question of the relative suppression of the (leading) 
UV renormalon contributions in RScl-- and RSch--invariant resummations 
would deserve a further study.
An additional uncertainty resides in the fact
that the residues, in contrast to the
renormalon pole locations, change and thus attain
unknown values when we go beyond
the large--${\beta}_0$ approximation.
For the UV renormalons, this uncertainty
shows up in an especially acute form \cite{VZ}.

The afore--mentioned works, however, suggest
strongly that, in the BjPSR case
$S(Q^2_{\rm ph}\!=\!3$--$5 {\rm GeV}^2)$, we should
preferably fix the value of $c_3$ in our,
ECH, and TPS--PMS resummation approximants
by using ${\rm IR}_1$ ($z_{\rm pole}\!=\!1/{\beta}_0$)
and not ${\rm UV}_1$ ($z_{\rm pole}\!=\!-1/{\beta}_0$)
information. The ${\rm IR}_1$ pole location can be transcribed
as $y_{\rm pole}\!=\!2$, where $y\!\equiv\!2 \beta_0 z$.
This corresponds to possible
renormalon--ambiguity contributions $\sim$$1/Q_{\rm ph}^2$
to the BjPSR observable which are nonperturbative.

We will present now an algorithm 
for adjusting approximately the value of $c_3$ 
in our approximant for the NNLO TPS (\ref{S2TPS}).
Briefly, it consists of the requirement that
$c_3$ must be adjusted in such a way that
the Borel transform of the approximant 
has the correct known location of the lowest
positive pole, where the latter location is obtained
by construction of Pad\'e approximants (PA's) of
the Borel transform.

A first idea would be to use simple Borel transforms.
We would first expand our approximant
(with a general yet unspecified $c_3$) in power series of
a coupling parameter, say 
$a_0\!\equiv\!a(\ln Q_0^2; c_2^{(0)},c_3^{(0)},...)$, 
up to a certain order $\sim\!a_0^{j+1}$ ($j \geq 3$), 
then obtain from this predicted $S_{[j]}$ TPS the corresponding 
${\cal B}_{[j]}(z)$ TPS (up to $\sim$$z^j$) of the simple Borel 
transform as schematically described by
\begin{eqnarray}
\sqrt{ {\cal A}_{{\tilde S}}^{[2/2]}(a_0;c_3) } = 
S_{[j]}^{\rm pr.}(a_0;c_3) & = & 
a_0 \left[ 1 + r_1 a_0 + r_2 a_0^2 + r_3^{\rm pr.}(c_3) a_0^3 +
\cdots + r_j^{\rm pr.}(c_3) a_0^j \right] \ ,
\label{Aansexp}
\\
\Rightarrow \quad 
{\cal B}_{[j]}^{\rm pr.}(z;c_3) & = & 1 + \frac{r_1}{1!} z + 
\frac{r_2}{2!} z^2 + \frac{r_3^{\rm pr.}(c_3)}{3!} z^3 +
\cdots + \frac{r_j^{\rm pr.}(c_3)}{j!} z^j \ .
\label{Bjpr}
\end{eqnarray}
The (approximate) pole structure of the simple Borel 
transform can be investigated by constructing various PA's
of its TPS (\ref{Bjpr}).
The requirement that the lowest positive pole
be at $y$($\equiv\!2 \beta_0 z$)$=2.0$ would then
give us predictions for $c_3$. However, this approach
is in practice seriously hampered, because
coefficients $r_k/k!$ of the simple
Borel transform ${\cal B}(z;c_3)$ depend
very much on the choice of the RScl ($Q_0^2$) and 
RSch ($c_2^{(0)}, c_3^{(0)}, \ldots$).
For example, if expanding our approximant
$\sqrt{{\cal A}_{S^2}(a_0;c_3)}$ up to $\sim\!a_0^4$
in an RSch with $c_2^{(0)}\!=\!c_2^{{\overline {\rm MS}}}$ 
and an arbitrary $c_3^{(0)}$, and keeping the
RScl $Q_0^2$ unchanged ($=\!Q_{\rm ph}^2$),
we reproduce in the BjPSR case the
first two coefficients $r_1$ and $r_2$ of (\ref{TPSBj}),
while the predicted $r_3$ in this RSch is
\begin{equation}
r_3^{\rm pr.} = 125.790... - \frac{c_3^{(0)}}{2} + c_3 \ .
\label{r3pr}
\end{equation}
The PA's $[2/1]$ or $[1/2]$ of the
corresponding simple Borel transform 
TPS ${\cal B}^{\rm pr.}_{[3]}(z)$ 
would therefore be functions of $(- c_3^{(0)}/2 + c_3)$,
and the requirement $y_{\rm pole}\!=\!2.0$
would at this level give us only a prediction
for $(- c_3^{(0)}/2 + c_3)$, not for $c_3$ itself.\footnote{
The $[1/1]$ PA of the simple Borel transform
is independent of $c_3$ and of $c_3^{(0)}$.
In the BjPSR case, in the ${\overline {\rm MS}}$ RSch and at
RScl $Q_0^2\!=\!3$ or $5 {\rm GeV}^2$, where $n_f\!=\!3$,
it predicts $y_{\rm pole}\!\approx\!1.6$.}
For example, working with ${\cal B}_{[3]}^{\rm pr.}(z)$
in the RSch with $c_3^{(0)}\!=\!0$
results in a prediction for $c_3$ that is
by about $10.5$ lower than the one when
$c_3^{(0)}\!=\!c_3^{{\overline {\rm MS}}}$($\approx\!21$) is used.
If using the ECH $a_{\rm ECH}(c_3)$
\cite{ECH} or TPS-PMS $S_{\rm PMS}(c_3)$
\cite{PMS} approximants instead of our approximant
(where $c_3$ is the arbitrary subleading parameter used
in $a_{\rm ECH}$ and $a_{\rm PMS}$ -- cf.~Appendix C), 
the corresponding prediction with $Q_0\!=\!Q_{\rm ph}$ is: 
$r_3^{(\rm pr.)} = 129.8998...\!+\!(-c_3^{(0)}\!+\!c_3)/2$.
Hence, also in the case of these approximants we end up 
with the same kind of problem of strong RSch--dependence 
($c_3^{(0)}$--dependence) of the predicted values of $c_3$.

Therefore, we will use a variant of the 
RScl-- and RSch--independent Borel 
transform $B(z)$ introduced by Grunberg \cite{BTinv}, who in turn
introduced it on the basis of the modified Borel transform 
of the authors of Ref.~\cite{BTmod}
\begin{equation}
S(Q^2_{\rm ph}) = \int_0^{\infty} dz 
\exp \left[ - {\rho}_1 (Q^2_{\rm ph}) z \right] B_{S}(z) \ .
\label{BT1}
\end{equation}
Here, ${\rho}_1$ is the first Stevenson's RScl and RSch invariant
(\ref{rho1}) of the observable $S$: 
\begin{equation}
{\rho}_1(Q^2_{\rm ph}) = - r_1(Q^2_{\rm ph}/Q_0^2) +
\beta_0 \ln \frac{Q_0^2}{{\tilde \Lambda}^2} =
\beta_0 \ln \frac{Q^2_{\rm ph}}{{\overline \Lambda}^2} \ ,
\label{BT2}
\end{equation}
where ${\tilde \Lambda}$ is the universal scale
appearing in the Stevenson equation (\ref{Stevenson}),
while ${\overline \Lambda}$ is a scale which depends
on the choice of the observable $S$. 
But ${\overline \Lambda}$ is independent
of RScl $Q_0$ and of RSch and even of the process momentum
$Q_{\rm ph}$. We note that ${\rho}_1(Q^2_{\rm ph})$
is, up to a constant $c$ (the latter is irrelevant
for the position of the poles of $B_S$), equal to
$1/a^{\rm (1-loop)}(Q^2_{\rm ph})$.
Thus, $B_S(z)$ of (\ref{BT1}) reduces to the simple
Borel transform, up to a factor $\exp(c z)$, if higher
than one--loop effects are ignored. The positions of the
poles of  $B_S(z)$ of (\ref{BT1}) are the same as those 
of the simple Borel transform. The coefficients of the
power expansion of $B_S(z)$ of (\ref{BT1}) are
RScl-- and RSch--invariant, in contrast to the case
of the simple Borel transform.
These invariant coefficients can be related with
coefficients $r_n$ of $S$ with relative ease 
in a specific RSch $c_k\!=\!c_1^k$
($k\!=\!2,3,4,\ldots$),
while keeping the RScl $Q_0^2$ unchanged
\begin{equation}
B_S(z) = (c_1 z)^{c_1 z} \exp( - r_1 z)
\sum_0^{\infty} 
\frac{ ( {\tilde r}_n\!-\!c_1 {\tilde r}_{n-1} ) }
{ {\Gamma}(n\!+\!1\!+\!c_1 z) } z^n 
\equiv (c_1 z)^{c_1 z} \overline{B}_S(z) \ .
\label{BT3}
\end{equation}
Here, ${\tilde r}_n$ is the coefficient at ${\tilde a}^{n+1}$
in the expansion of $S$ in powers of
${\tilde a}\!\equiv\!a(\ln Q_0^2;c_1^2,c_1^3,c_1^4,...)$,
and by definition
${\tilde r}_{-1}\!=\!0$, ${\tilde r}_{0}\!=\!1$.
In (\ref{BT3}), we introduced the modified
RScl-- and RSch--invariant Borel transform 
$\overline{B}_S(z)$, by extracting the
factor $(c_1 z)^{c_1 z}$ whose behavior
at $z\!\to\!0$ may be problematic for
PA's to deal with.\footnote{
Grunberg's \cite{BTinv} Borel transform 
${\widetilde B}^{({\rm Gr.})}$ was chosen by convention as: 
${\widetilde B}^{({\rm Gr.})}(z)\!=\!{\Gamma}(1\!+\!c_1 z) 
\exp(c_1 z) \overline{B}(z)$.
In this way, 
${\widetilde B}^{({\rm Gr.})}(z)\approx B(z) \sqrt{2 \pi c_1 z}$
when $z\!\to\!\infty$, and the coefficients of the power
expansion of ${\widetilde B}^{({\rm Gr.})}$ in $z$ depend only
on the RScl-- and RSch--invariants $\rho_j$ (no dependence
on $c_1$ and on $\Gamma$--function--related constants).
We decided not to follow this convention, primarily 
since ${\Gamma}(1\!+\!c_1 z)$ introduces spurious poles
on the negative axis, the one closest to the origin being 
$y$($\equiv\!2 {\beta}_0 z$)$\approx\!-2.53$.
Such spurious poles not far away from the origin can 
significantly limit the PA's ability to
locate correctly the leading IR renormalon
pole ($y_{\rm pole}\!\approx\!2.$).} 
The obtained coefficients of the power expansion
of $\overline{B}_S(z)$ are explicitly RScl-- and RSch--invariant,
depending only on the invariants ${\rho}_j$ ($j\geq 2$),
on $c_1$ and on some universal constants.

We will now calculate the invariant Borel transform
$\overline{B}_{{\sqrt {\cal A}}}$ of our approximant.
The coefficients ${\tilde r}_k$ as predicted by our approximant
(\ref{Aans}) $\sqrt{{\cal A}_{S^2}}(c_3)$ 
are functions of the only unknown
$c_3$ [${\tilde r}_k\!=\!{\tilde r}_k(c_3)$, $k \geq 3$],
They can be obtained as coefficients of the power expansion
of $\sqrt{{\cal A}_{S^2}}(c_3)$ in powers of ${\tilde a}$.
Looking back at the form (\ref{Aans}) of our approximant,
such a power expansion requires first the separate
expansions of $a_1\!=\!a(Q_1^2;c_2^{(1)},c_3,0,\ldots)$
and of $a_2\!=\!a(Q_2^2;c_2^{(2)},c_3,0,\ldots)$
in powers of ${\tilde a}$. The latter expansions
can be read off Eq.~(\ref{aexpina0}), up to $\sim\!{\tilde a}^5$
(there: $a \mapsto a_1$ or $a_2$; and $a_0 \mapsto {\tilde a}$.) 
In fact, we carried
out the latter expansion up to $\sim\!{\tilde a}^8$
(with the help of Mathematica), which allowed us
to write the approximant $\sqrt{{\cal A}_{S^2}}(c_3)$
up to $\sim\!{\tilde a}^7$. This in turn leads us to
obtain the invariant Borel transform $B_{\sqrt{\cal A}}(z)$
up to $\sim\!z^6$, according to (\ref{BT3}),
and allows us to construct PA's of the Borel transform
of as high order as $[3/3]$, $[2/4]$, $[5/1]$.
The coefficients starting at $z^3$ are predictions
of the approximant and are $c_3$--dependent:
${\overline B}_S(z)\!=\!1\!+\!{\bar b}_1 z\!+\!{\bar b}_2 
z^2\!+\!{\bar b}_3(c_3) z^3\!+\cdots$, with 
${\bar b}_1\!\approx\!-0.7516$,
${\bar b}_2\!\approx\!0.4209$,
${\bar b}_3(c_3)\!\approx\!(-2.664\!+\!0.1667 c_3)$, etc.
Construction of various PA's
of that Borel transform and requirement that
the smallest positive pole equal
$y_{\rm pole}$($=\!2 \beta_0 z_{\rm pole}$)$=2.0$
gives us predictions for $c_3$ which are
listed for the described case in the second column of
Table \ref{tabl1}.
In the column we included values of $c_3$
with small nonzero imaginary parts and 
${\rm Re} (c_3)\!\approx\!10$--$12$, since for
such values the ${\rm PA}_{\overline{B}}$'s 
and the TPS of $\overline{B}$
are almost real, with imaginary
parts less than one per cent of the real part
for $y < 1.9$. In the latter cases the real part of $c_3$
may be regarded as the suggested value. The actual value of $c_3$
must be exactly real, but since a specific PA predicts only an approximate
value of $c_3$, this latter value is not necessarily exactly real. 
\begin{table}[ht]
\par
\begin{center}
\begin{tabular}{l c  c  c}
${\rm PA}_{\overline{B}}$  & 
$c_3$ ($\sqrt{{\cal A}_{S^2}}$) &
$c_3$ (ECH) & $c_3$ (TPS--PMS) \\
\hline \hline
$[2/1]$ & 21.7 & 35.1 & 35.1 \\
\hline
$[3/1]$ & 13.7 & 19.5 & 19.0 \\
\hline
$[4/1]$ & 11.1 & 14.4 & 13.1 \\
\hline
$[5/1]$ & 9.3  & 11.2 & 8.8 \\
\hline\hline
$[1/2]$ & 12.8 & 17.3 & 17.3 \\
\hline
$[2/2]$ & 12.4 & 16.9 & 16.2 \\
\hline
$[3/2]$ & $11.7\!\pm\!3.4 {\rm i}$ &
$15.8\!\pm\!6.4 {\rm i}$ &
$15.4\!\pm\!7.4 {\rm i}$ \\
\hline
$[4/2]$ & $10.3\!\pm\!2.8 {\rm i}$ &
$12.9\!\pm\!5.1 {\rm i}$ &
$11.6\!\pm\!6.8 {\rm i}$ \\
\hline\hline
$[1/3]$ & 12.4 & 16.9 & 16.2 \\
\hline
$[2/3]$ & 12.9 & 17.4 & $18.3\!\pm\!0.8 {\rm i}$ \\
\hline
$[3/3]$ & 
$10.6\!\pm\!2.9 {\rm i}$ &
$13.6\!\pm\!5.5 {\rm i}$ &
$12.6\!\pm\!7.0 {\rm i}$ \\
\hline\hline
average & $\approx 12.5$ & $\approx 17.$ & $\approx 16.$
\end{tabular}
\end{center}
\caption {\footnotesize Predictions for $c_3$ in our, ECH and
TPS--PMS approximants, using various PA's of the invariant
Borel transform $\overline{B}(z)$ of the approximants
and demanding that the lowest positive pole be at
$z_{\rm pole}\!=\!1/{\beta}_0$ ($=4/9$). The higher order
parameters $c_k^{(j)}$ ($k \geq 4$, $j\!=\!1,2$) in our approximant,
and $c_k$ ($k \geq 4$) in ECH and TPS--PMS, were all set
equal to zero.}
\label{tabl1}
\end{table}
We did not include some other solutions which differ
a lot from those given in the column.
Predictions of ${\rm PA}_{\overline{B}}$'s of the
intermediate orders ([3/1], [4/1], [2/2], [3/2], [1/3], [2/3])
give us the average value $c_3\!\approx\!12.5$ which we will adopt. 
The prediction by 
PA $[2/1]$ differs from most of the other predictions, 
apparently because $[2/1]$ is of low order. Predictions by
the highest PA's ($[5/1]$, $[4/2]$, $[3/3]$) also differ from 
the average. The reason for this lies probably in the fact
that these PA's contain information on many higher order 
coefficients ${\tilde r}_n$ ($n\!=\!3,4,5,6$) which are not 
contained in the TPS $S_{[2]}$ on which the approximant 
$\sqrt{{\cal A}_{S^2}}$ is based. In addition, these high
order PA's are implicitly dependent on the high order
parameters $c_k^{(1)}$ and $c_k^{(2)}$ ($k\!=\!4,5,6,7$)
which were here simply set equal to zero (we will come back
to this point later in Section V). 

Completely analogous considerations produce
the values of $c_3$ parameter in the
ECH and TPS-PMS approximants. For details on
the ECH and TPS--PMS methods, when applied to the
NNLO TPS $S_{[2]}$ (\ref{S2TPS}), we refer to
Appendix C. Also in this case, we make
for the corresponding $\beta$--functions the simple
TPS choice:
ECH RSch$= ({\rho}_2, c_3, 0, ...)$;
TPS-PMS RSch$=(3 {\rho}_2/2, c_3, 0, ...)$.
The obtained predictions for $c_3$ for these
approximants are included in Table \ref{tabl1}.
Again, PA $[2/1]$ and
the highest order PA's appear to give unreliable predictions.
On the basis of the predictions of ${\rm PA}_{\overline{B}}$'s
of intermediate order,
we will adopt the value $c_3\!=\!17$ for the
ECH case, and $c_3\!=\!16$ for the TPS--PMS case.
The actual values of $c_3$ must be exactly real.

In fact, we can apply this method of determining
the $c_3$--parameter of our approximant (and of
ECH and TPS--PMS approximants) to
any QCD observable given at the NNLO and whose
leading IR renormalon pole is known 
via large--${\beta}_0$ considerations. The method,
however, is well motivated only if there are indications 
that the leading IR renormalon contributions
to the observable are larger than those of the
leading UV renormalon. We wish to stress that
our approximant, as well as the ECH and TPS--PMS
approximants, are completely independent of the
original choice of the RScl and RSch in the TPS of the 
observable, because the parameter $c_3$ is 
RScl-- and RSch--invariant since it is 
determined by using the RScl-- and RSch--invariant Borel transform
${\overline B}(z)$.

A few remarks about the multiplicity of the discussed
${\rm IR}_1$ pole are in order. The simple
Borel transform $\sum r_k z^k/k!$ of $S(Q^2_{\rm ph})$
behaves near $z_{\rm pole}$ ($\!=\!1/{\beta}_0$) as 
$\sim\!1/(z_{\rm pole}\!-\!z)^\kappa$ where 
the multiplicity is \cite{Mueller,Grunberg2,VZ} 
$\kappa\!=\!1\!+\!({\beta}_1/{\beta}_0) z_{\rm pole}\!+\!({\gamma}/
{\beta_0})$, and ${\gamma}$ is the one--loop anomalous
dimension of the corresponding two--dimensional 
operator appearing in the Operator Product Expansion for $S$
(usually ${\gamma} \geq 0$).
On the other hand, the RScl-- and RSch--invariant Borel transform
(\ref{BT3}) behaves near $z_{\rm pole}$ with the
simpler pole multiplicity \cite{BTmod}
$\kappa\!=\!1\!+\!({\gamma}/{\beta_0})$.
To our knowledge, the anomalous dimension ${\gamma}$
is not known in this case. However, in the case of the
Adler function (logarithmic derivative of the correlation
function of quark current operators), the one--loop anomalous
dimension of the four--dimensional operator corresponding
to the lowest IR renormalon pole there 
($z_{\rm pole}\!=\!2/{\beta}_0$)
is known \cite{Mueller,CDJ} to be $\gamma\!=\!0$.
If $\gamma\!=\!0$ also in the BjPSR case, then
the RScl-- and RSch--invariant Borel transform (\ref{BT1})--(\ref{BT3})
has $\kappa\!=\!1$, i.e., the leading IR renormalon
pole is a simple pole, in contrast to the simple
Borel transform where $\kappa$ is noninteger.
In such a case, we may have an additional incentive
to use, instead of the simple Borel transform, the 
invariant Borel transform (\ref{BT1})--(\ref{BT3})
in conjunction with the afore--described
PA's of Table \ref{tabl1}. Namely, PA's are very
good at discerning the location of a pole if
such a pole is simple, and are somewhat less
successful in this job if the pole is multiple
or with noninteger multiplicity. 

\section{BJPSR: predictions for the coupling parameter}

Now that we have fixed the values
of the $c_3$--parameter in the approximants
$\sqrt{{\cal A}_{S^2}}(a_0;c_3)$,
ECH and TPS--PMS,
the only adjustable parameter in them is the numerical value of 
$a_0\!\equiv\!{\alpha}_s^{{\overline {\rm MS}}}(Q^2_{\rm ph})/\pi$, 
at such $Q^2_{\rm ph}$ where three flavors
are assumed active, e.g. at $Q^2_{\rm ph}\!=\!3$ or $5 \ {\rm GeV}^2$. 
This $a_0$ can be obtained by requiring that
it should reproduce the experimental values for $S(Q^2_{\rm ph})$
of (\ref{BjPSR1}). The questions connected with
the extraction of the values of the BjPSR integral 
(\ref{BjPSR1}) from the measured polarized structure
functions are at present not quite settled. One source of
the uncertainty arises from the fact that these structure
functions have not been measured at small values of
$x_{\rm Bj}$ and that, therefore, a theoretical extrapolation
to such small $x_{\rm Bj}$--values is needed. 
The authors of \cite{EK,EGKS}
used the small--$x_{\rm Bj}$ extrapolation as suggested by
the Regge theory, the assumption made also by various
experimentalist groups before 1997. The values thus obtained
by \cite{EK,EGKS}, on the basis of measurements at SLAC 
and CERN before 1997, are
\begin{equation}
({\rm{Regge}}): \quad 
\frac{1}{6} |g_A| \left[ 1 - S(Q^2_{\rm ph}\!=\!3 {\rm GeV}^2) 
\right] = 0.164 \pm 0.011 \ .
\label{Regge}
\end{equation}  
On the other hand, the authors of \cite{ABFR} 
used a small--$x_{\rm Bj}$ extrapolation based on 
the NLO version of the DGLAP equations (pQCD)
as opposed to the Regge extrapolation
(cf.~also Ref.~\cite{BFR}).
This leads to higher values and
larger uncertainties of the BjPSR integral.
The values extracted in this way by \cite{ABFR} (their Table 4), 
based on SLAC data, are
\begin{equation}
(II): \quad \frac{1}{6} |g_A| \left[ 1 - S(Q^2_{\rm ph}\!=\!3 {\rm GeV}^2) 
\right] = 0.177 \pm 0.018 \ .
\label{BjdatII}
\end{equation}
Furthermore, most of the experimentalist groups have adopted,
since 1997, similar NLO pQCD approaches to the small-$x_{\rm Bj}$
extrapolation, e.g. SMC Collaboration \cite{1997SMC} at CERN,
E154 \cite{1997E154} and E155 \cite{2000E155}
Collaborations at SLAC. The most recent and updated
measurements of the polarized structure functions
are those of Ref.~\cite{2000E155}. Their combined value
of the BjPSR--integral at $Q_{\rm ph}^2\!=\!5 \ {\rm GeV}^2$ is
\begin{equation}
(I): \quad \frac{1}{6} |g_A| \left[ 1 - S(Q^2_{\rm ph}\!=\!5 {\rm GeV}^2) 
\right] = 0.176 \pm 0.008 \ .
\label{BjdatI}
\end{equation}
Apart from the problem of the small--$x_{\rm Bj}$ extrapolation,
there is a problem of accounting for nuclear effects.
Since the extraction of the $g_1^{(n)}$ structure function
is based on the measurements of the structure functions
of the deuteron and ${}^3{\rm He}$, nuclear
effects have to be taken into consideration.
The (multiplicative) effects due to the nuclear 
wavefunction have been taken into
account in (\ref{BjdatI}) and (\ref{BjdatII}).
However, recently the authors of 
\cite{SchmidtYang}
argued that additional nuclear effects, originating from
spin--one isosinglet 6--quark clusters in deuteron and helium
(which include the shadowing, EMC and Fermi motion effects), 
affect the extracted values of the neutron structure function
$g_1^{(n)}$ in such a way that the value of the BjPSR
integral increases by about $10\%$.
This would then change the E155 values of (\ref{BjdatI}) to
\begin{equation}
(I^{\prime}): \quad \
\frac{1}{6} |g_A| \left[ 1 - S(Q^2_{\rm ph}\!=\!5 {\rm GeV}^2) 
\right] = 0.193 \pm 0.009 \ .
\label{BjdatIpr}
\end{equation}  
The values of the (\ref{BjdatII}), at $Q_{\rm ph}^2\!=\!3 {\rm GeV}^2$
would be increased to about $0.195 \pm 0.020$. We will
not consider this case ${\rm II}^{\prime}$
and case ${\rm I}^{\prime}$ (\ref{BjdatIpr})
for the time being, but will briefly return to them in Section VI.

In the following we will extract the values of 
$\alpha_s^{{\overline {\rm MS}}}(Q^2_{\rm ph})$ from
the BjPSR--integral values (\ref{BjdatI}) and
(\ref{BjdatII}), and will simply
denote the corresponding cases as I and II, respectively.

If we insert the value (\ref{gA}) for $|g_A|$
into (\ref{BjdatI}) and (\ref{BjdatII}), we obtain 
\begin{eqnarray}
(I): \quad  S(Q^2_{\rm ph}\!=\!5 \ {\rm GeV}^2) &=& 0.167 \pm 0.038 \ ,
\label{SI}
\\
(II): \quad  S(Q^2_{\rm ph}\!=\!3 \ {\rm GeV}^2) &=& 0.162 \pm 0.085 \ .
\label{SII}
\end{eqnarray}
The present small uncertainty in the value of $|g_A|$ (\ref{gA})
practically does not contribute to the uncertainties of
$S(Q^2_{\rm ph})$ in (\ref{SI})--(\ref{SII}). 

Our approximant gives, for example, for 
$a_0\!\equiv\!a(\ln 3 {\rm GeV}^2; 
c_2^{{\overline {\rm MS}}}, c_3^{{\overline {\rm MS}}}, 
0, \ldots)\!=\!0.09$ 
[$\Leftrightarrow \ 
\alpha_s^{{\overline {\rm MS}}}(Q^2\!=\!3 {\rm GeV}^2) 
\approx 0.283$]
the value $0.1585$, which is not far from the middle values
in (\ref{SI})--(\ref{SII}). Varying $a_0$ in our approximant
(with $c_3\!=\!12.5$) in such a way that the middle and the 
end--point values of the right--hand side of (\ref{SI}) 
or (\ref{SII}) are reproduced then results in the following
predictions for $\alpha_s$ 
(in ${\overline {\rm MS}}$ RSch):
\begin{equation}
\alpha_s^{{\overline {\rm MS}}}(Q^2\!=\!5 \ {\rm GeV}^2) = 
0.2894^{+0.0238}_{-0.0345}  \ (I) \ ; \quad 
\alpha_s^{{\overline {\rm MS}}}(Q^2\!=\!3 \ {\rm GeV}^2) = 
0.2855^{+0.0450}_{-0.1024}  \ (II) \ .
\label{alph3}
\end{equation}
We then evolved these predicted values via four--loop
RGE (\ref{RGE}) to $Q^2\!=\!M_Z^2$, using the values of
the four--loop coefficient $c_3(n_f)$ in
the ${\overline {\rm MS}}$ RSch \cite{RVL}
and the corresponding three--loop matching
conditions \cite{Chetyrkinetal} for the flavor
thresholds. We used the matching at 
$\mu(n_f) = \kappa m_q(n_f)$ with the choice
$\kappa\!=\!2$, where $m_q(n_f)$
is the running quark mass $m_q(m_q)$
of the $n_f$'th flavor and $\mu(n_f)$
is defined as the scale above which $n_f$ 
flavors are active.\footnote{ 
If increasing $\kappa$ from $1.8$ to $3$ in case I, 
the predictions for the central, upper, lower values 
of $\alpha_s(M_Z^2)$ decrease by $0.12 \%$, 
$0.15 \%$, $0.09 \%$, respectively;
increasing $\kappa$ from $1.5$ to $3$ in case II, 
the respective numbers are
$0.12 \%$, $0.17 \%$, $0.03 \%$.
We assumed $m_c(m_c)\!=\!1.25$ GeV and $m_b(m_b)\!=\!4.25$ GeV.
\label{thresh}
}
The resulting predictions for $\alpha_s(M_Z^2)$ are
\begin{equation}
\alpha_s^{{\overline {\rm MS}}}(M_Z^2) = 
0.1196^{+0.0035}_{-0.0059} \ (I) \ ; \quad
0.1135^{+0.0058}_{-0.0196} \ (II) \ . 
\label{alphMZ}
\end{equation}
In Table \ref{tabl2}, we give the values of
${\alpha}^{\overline {\rm MS}}_s$ as 
predicted from the BjPSR data
(\ref{SI}) and (\ref{SII})
by our approximant (with $c_3\!=\!12.5$), by
the ECH (with $c_3\!=\!17$), and by the TPS--PMS 
(with $c_3\!=\!16$). For comparison, we include predictions
of these three approximants when $c_3$ in them is set equal to zero,
i.e., for the case when the location of the leading IR renormalon
(${\rm IR}_1$) pole in these approximants is not correct. 
\begin{table}[ht]
\par
\begin{center}
\begin{tabular}{l| c c| c c}
approximant & 
$\alpha_s( 5 {\rm GeV}^2)$: (I) & 
$\alpha_s( 3 {\rm GeV}^2)$: (II) &
$\alpha_s(M_Z^2)$: (I) & $\alpha_s(M_Z^2)$: (II) \\
\hline \hline
NNLO TPS & 
$0.3287^{+0.0465}_{-0.0530}$ & 
$0.3221^{+0.0989}_{-0.1341}$ & 
$0.1252^{+0.0055}_{-0.0078}$ &
$0.1183^{+0.0095}_{-0.0232}$ \\
\hline
${\rm N}^3{\rm LO}$ TPS ($r_3\!=\!128.05$) & 
$0.3121^{+0.0393}_{-0.0464}$ &
$0.3065^{+0.0823}_{-0.1215}$ &
$0.1230^{+0.0050}_{-0.0073}$ &
$0.1163^{+0.0089}_{-0.0219}$ \\
\hline \hline
$[1/2]_S$ (NNLO) &
$0.3054^{+0.0339}_{-0.0426}$ &
$0.3003^{+0.0693}_{-0.1155}$ &
$0.1220^{+0.0046}_{-0.0069}$ &
$0.1155^{+0.0079}_{-0.0212}$ \\
\hline
$[2/1]_S$ (NNLO) &
$0.3006^{+0.0316}_{-0.0404}$ &
$0.2959^{+0.0637}_{-0.1118}$ &
$0.1213^{+0.0044}_{-0.0066}$ &
$0.1149^{+0.0075}_{-0.0208}$ \\
\hline
$\sqrt{[2/2]_{S^2}}$ (NNLO) & 
$0.2937^{+0.0271}_{-0.0369}$ &
$0.2895^{+0.0533}_{-0.1061}$ &
$0.1203^{+0.0039}_{-0.0063}$ &
$0.1140^{+0.0066}_{-0.0200}$ \\
\hline\hline
$[2/2]_S$ (${\rm N}^3{\rm LO}$, $r_3\!=\!128.05$)  & 
$0.2944^{+0.0282}_{-0.0375}$ &
$0.2901^{+0.0561}_{-0.1067}$ &
$0.1204^{+0.0040}_{-0.0063}$ &
$0.1141^{+0.0069}_{-0.0201}$ \\
\hline \hline
TPS--PMS (NNLO, $c_3\!=\!16.$) & 
$0.2907^{+0.0259}_{-0.0354}$ &
$0.2867^{+ {\rm ?}}_{-0.1035}$ &
$0.1198^{+0.0038}_{-0.0060}$ &
$0.1136^{+ {\rm ?}}_{-0.0197}$ \\
\hline
ECH (NNLO, $c_3\!=\!17.$) &
$0.2898^{+0.0244}_{-0.0348}$ &
$0.2859^{+0.0468}_{-0.1028}$ &
$0.1196^{+0.0037}_{-0.0059}$ &
$0.1135^{+0.0060}_{-0.0196}$ \\
\hline\hline
$\sqrt{ {\cal A}_{S^2}^{[2/2]} }$ (NNLO, $c_3\!=\!12.5$) & 
$0.2894^{+0.0238}_{-0.0345}$ &
$0.2855^{+0.0450}_{-0.1024}$ &
$0.1196^{+0.0035}_{-0.0059}$ &
$0.1135^{+0.0058}_{-0.0196}$ \\
\hline\hline
TPS--PMS (NNLO, $c_3\!=\!0.$) & 
$0.2957^{+0.0296}_{-0.0380}$ &
$0.2913^{+ {\rm ?}}_{-0.1077}$ &
$0.1206^{+0.0042}_{-0.0064}$ &
$0.1143^{+ {\rm ?}}_{-0.0203}$ \\
\hline
ECH (NNLO, $c_3\!=\!0.$) &
$0.2947^{+0.0273}_{-0.0373}$ &
$0.2904^{+0.0537}_{-0.1068}$ &
$0.1204^{+0.0039}_{-0.0062}$ &
$0.1142^{+0.0066}_{-0.0202}$ \\
\hline
$\sqrt{ {\cal A}_{S^2}^{[2/2]} }$ (NNLO, $c_3\!=\!0.$) & 
$0.2960^{+0.0278}_{-0.0378}$ &
$0.2916^{+0.0545}_{-0.1078}$ &
$0.1206^{+0.0040}_{-0.0063}$ &
$0.1143^{+0.0067}_{-0.0202}$ 
\end{tabular}
\end{center}
\caption {\footnotesize Predictions for 
${\alpha}_s^{{\overline {\rm MS}}}$,
derived from various resummation approximants to the BjPSR at
$Q^2_{\rm photon}\!=\!5 {\rm GeV}^2$, $3 \ {\rm GeV}^2$. 
Predictions for the case
I (\ref{BjdatI}) and II (\ref{BjdatII}) are given in parallel.}
\label{tabl2}
\end{table}
Given are always three predictions for $\alpha_s$,
corresponding to the three values of $S$ (\ref{SI})
for case I, and (\ref{SII}) for case II.
In addition, predictions of the following approximants
are included in Table \ref{tabl2}: 
TPS $S_{[2]}$ (\ref{TPSBj}) (NNLO TPS);
TPS $S_{[3]}$ with $r_3\!=\!128.05$ (${\rm N}^3{\rm LO}$ TPS);
off--diagonal Pad\'e approximants (PA's) $[1/2]_S$ and
$[2/1]_S$, both based solely on 
the NNLO TPS $S_{[2]}$ (\ref{TPSBj});
square root of the diagonal PA (dPA) $[2/2]_{S^2}$,
which is based solely on the NNLO TPS (\ref{TPSBj});
$[2/2]_S$ is the dPA constructed on the basis of
the ${\rm N}^3{\rm LO}$ TPS $S_{[3]}$ with $r_3\!=\!128.05$.
For $[2/2]_S$ and ${\rm N}^3{\rm LO}$ TPS we chose the latter value 
of $r_3$ (in ${\overline {\rm MS}}$, at RScl 
$Q_0^2\!=\!Q_{\rm ph}^2$, $n_f\!=\!3$) because then the $[1/2]$ PA
of the invariant Borel transform $\overline{B}_S$ (\ref{BT3})
predicts the ${\rm IR}_1$ pole
$y_{\rm pole}\!=\!2.0$. 
We wrote in Table \ref{tabl2}
numbers with four digits in order to facilitate a clearer
comparison of predictions of various methods.
 
{}From Table \ref{tabl2} we see that
the values of $\alpha_s^{\overline {\rm MS}}(M_Z^2)$ predicted by
various approximants differ significantly from each other.
Addition of the ${\rm N}^3 {\rm LO}$ term in the TPS 
decreases the central value of 
$\alpha_s^{{\overline {\rm MS}}}(M_Z^2)$ 
by $0.0022$ ($0.0020$ in case II),
and application of the NNLO dPA approximant $[2/2]_{S^2}^{1/2}$ 
decreases this value by a further $0.0027$ ($0.0023$). 
Our approximant $\sqrt{{\cal A}_{S^2}}(c_3\!=\!12.5)$, which
is an RScl-- and RSch--invariant extension of the method of
the dPA $[2/2]_{S^2}^{1/2}$, decreases the central
$\alpha_s^{{\overline {\rm MS}}}(M_Z^2)$ by
a further amount of $0.0007$ ($0.0005$).
Predictions of the ECH and TPS--PMS methods are
very close to those of our method if the value of $c_3$
in them is adjusted in the afore--described way. However, 
predictions of these two and of our method increase and 
come closer to the predictions of the NNLO dPA once
we simply set in these approximants $c_3\!=\!0$, thus 
abandoning the requirement of the correct location
of the ${\rm IR}_1$ pole.
The predictions of the ${\rm N}^3{\rm LO}$ dPA $[2/2]_S$
are almost identical with those of the NNLO dPA.
All the PA resummations were carried out with the
RScl $Q_0^2\!=\!Q^2_{\rm ph}$ ($n_f\!=\!3$)
and in ${\overline {\rm MS}}$ RSch, and their predictions
would change somewhat if the RScl and RSch were changed
-- in contrast to the presented predictions of
$\sqrt{{\cal A}_{S^2}}$, ECH and TPS--PMS.

We wish to point out that the 
${\alpha}_s^{{\overline {\rm MS}}}$--predictions
for the case II (\ref{BjdatII}) were already presented
in the short version \cite{brief}. However, they were somewhat
lower there [the central values of 
${\alpha_s}^{{\overline {\rm MS}}}(M^2_Z)$ were
lower by about $0.0009$--$0.0011$] -- because the value
of the $\beta$--decay parameter $|g_A|$ there
was taken from the Particle Data Book of 1994
$|g_A|\!=\!1.257 (\pm 0.2\%)$
(used also in \cite{ABFR}), while the value used here (\ref{gA})
is the updated value based on \cite{PDG}.
 
\begin{figure}[ht]
\centering\epsfig{file=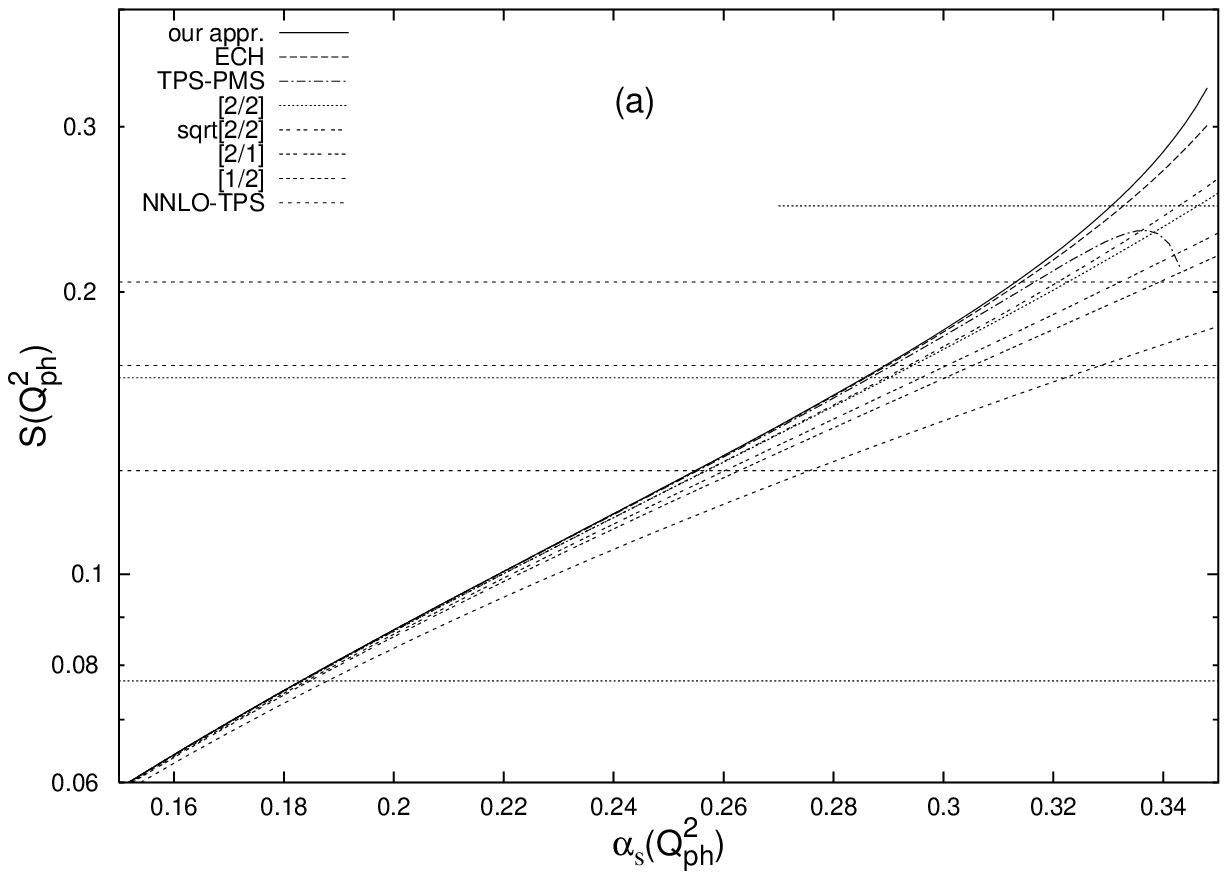,height=9.5cm}
\centering\epsfig{file=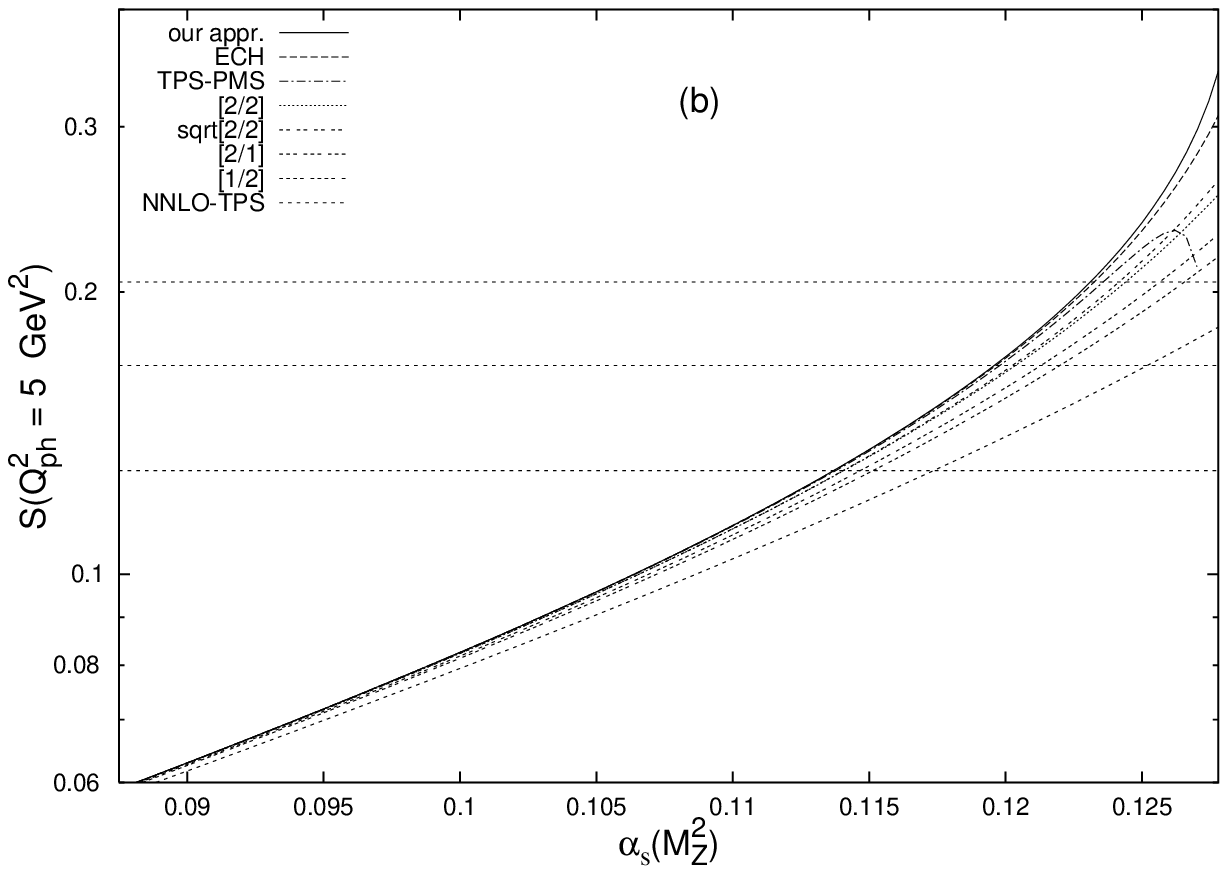,height=9.5cm}
\vspace{0.3cm}
\caption{Predictions of various approximants: 
(a) for $S(Q_{\rm ph}^2)$ as functions of 
${\alpha}^{{\overline {\rm MS}}}_s(Q_{\rm ph}^2)$ when $n_f\!=\!3$;
(b) for $S(Q_{\rm ph}^2\!=\!5{\rm GeV}^2)$ as functions of 
${\alpha}^{{\overline {\rm MS}}}_s(M_Z^2)$.
The values of the $c_3$--parameter in our approximant ($c_3\!=\!12.5$), 
ECH ($c_3\!=\!17.$)
and TPS--PMS ($c_3\!=\!16.$) have been adjusted to ensure
the correct location of the leading IR renormalon pole.
The experimental bounds $S_{\rm min}$, $S_{\rm max}$ and $S_{\rm mid}$ 
are indicated as dashed horizontal lines for case I (\ref{SI})
($Q_{\rm ph}^2\!=\!5 {\rm GeV}^2$)
and dotted horizontal lines for case II (\ref{SII})
($Q_{\rm ph}^2\!=\!3 {\rm GeV}^2$).} 
\label{fig1Bjf}
\end{figure}

In Fig.~\ref{fig1Bjf}(a)
we present various
approximants for $S(Q_{\rm ph}^2)$ as functions of 
${\alpha}^{{\overline {\rm MS}}}_s(Q_{\rm ph}^2)$
($n_f\!=\!3$, e.g. $Q_{\rm ph}^2\!=\!3$ or $5 {\rm GeV}^2$), 
and in Fig.~\ref{fig1Bjf}(b) the approximants for
$S(5{\rm GeV}^2)$ as functions of 
${\alpha}^{{\overline {\rm MS}}}_s(M_Z^2)$.
There is one peculiarity of the (NNLO) TPS--PMS method,
as seen also in Figs.~\ref{fig1Bjf} -- for high
values of observable $S$ this method does not
give solutions. This is so because the polynomial form 
of the (NNLO) TPS--PMS $S^{\rm PMS}$ [see Eq.~(\ref{SPMS})]
is bounded from above by $S_{\rm max}^{\rm PMS} =
(2/3)^{3/2} {\rho_2}^{-1/2}$ which, in the 
considered case ($\rho_2 = 5.476$), is equal
to $0.233$ which is below $S_{\rm max}\!=\!0.247$ in case II
(cf. Appendix C for more details).
This is also indicated in Table \ref{tabl2}.
 
We wish to emphasize one aspect that makes the approximant
$\sqrt{{\cal A}_{S^2}}$ 
conceptually quite different from the dPA $[2/2]_S$. Although
both approximants incorporate information about the location
of the ${\rm IR}_1$ pole ($y_{\rm pole}\!=\!2$),
they do it in two very different ways. The dPA
$[2/2]_S$ is constructed on the basis of the ${\rm N}^3{\rm LO}$
TPS with $r_3\!=\!128.05$, where only this latter coefficient
contains approximate information on the pole's location. 
So this dPA is a pure ${\rm N}^3{\rm LO}$--construction
and is RSch-- and even RScl--dependent (weakly).
The approximant $\sqrt{{\cal A}_{S^2}}$ 
is constructed on the basis of the NNLO
TPS. It is a RScl-- and $c_2^{(0)}$--independent 
NNLO--construction, and the
correct ${\rm IR}_1$ pole location is obtained by
the adjustment of the $c_3$--parameter within the approximant.
As argued previously [cf.~2nd paragraph after (\ref{PMSc3})],
the $c_3$--dependence in $\sqrt{{\cal A}_{S^2}}(c_3)$ is closely
related with the sensitivity of the approximant to the
details of the RGE evolution, 
and the latter details are the more important the more
nonperturbative the observable is. So it seems
very natural that it is the intrinsic $c_3$--parameter in 
$\sqrt{{\cal A}_{S^2}}(c_3)$ that parametrizes the (nonperturbative)
${\rm IR}_1$ pole location, and at the same time
it makes the approximant fully RSch--independent. 
The same is true for the
ECH and the TPS--PMS approximants. 

On the other hand, it would be an ambiguous approach
to implement this kind of $c_3$--fixing
in the NNLO PA methods ($[1/2]_S$, $[2/1]_S$,
$[2/2]_{S^2}^{1/2}$) -- because these resummations
depend in addition on the leading RSch--parameter $c_2$
($\Leftrightarrow c_2^{(0)}$) and even on the RScl $Q_0^2$.
Therefore, it may not be so surprising
that the results of our method, ECH,
and TPS--PMS, with the mentioned $c_3$--fixing,
all give predictions that are clustered closely
together and are significantly distanced
from the predictions of (d)PA's.

There is another theoretical aspect which indicates that
the predictions of the (NNLO) approximant 
$\sqrt{{\cal A}_{S^2}}$ should in general
be better than those of the
(NNLO) dPA $[2/2]^{1/2}_{S^2}$.
Namely, the latter dPA is
just a one--loop approximation to our approximant.
More specifically, dPA $[2/2]^{1/2}_{S^2}$ 
is like ansatz (\ref{Aans}),
but each $a_j\!\equiv\!a(\ln Q_j^2;c_2^{(j)},c_3,\ldots)$
is replaced by the coupling parameter 
$a^{ ({\rm 1-l.}) }(\ln {\bar Q}_j^2)$ evolved from
the RScl $Q_0^2$ to a ${\bar Q}_j^2$ by the one--loop RGE
in the original (${\overline {\rm MS}}$) RSch.
This follows from considerations in \cite{GC1,CK},
and can also be checked directly as indicated in
the paragraph after Eqs.~(\ref{Aold})--(\ref{param2old}). 
The dPA $[2/2]^{1/2}_{S^2}$
possesses residual RScl--dependence, and RSch--dependence, 
the unphysical properties not shared by the 
true (unknown) sum.
The approximant $\sqrt{{\cal A}_{S^2}}$, however,
possesses RScl-- and RSch--independence,
and is thus better suited to bring us closer to
the true sum.

On the other hand, when compared with the structure
of the ECH and TPS--PMS approximants,
$\sqrt{{\cal A}_{S^2}}$
possesses a theoretically favorable ``PA--type'' feature that the
other two methods don't have: It represents an efficient quasianalytic
continuation of the NNLO TPS $S_{[2]}$ from the
perturbative (small--$a$) to the nonperturbative
(large--$a$) regime. This is so because 
$\sqrt{{\cal A}_{S^2}}$
is related with the mentioned dPA method
$[2/2]_{S^2}^{1/2}$ (see above). 
The ECH and the TPS--PMS approximants
don't possess this strong type of 
mechanism of quasianalytic continuation,
because they don't go beyond the polynomial TPS structure
of the original TPS $S_{[2]}$. 
These two approximants do possess, however, a weaker type 
of quasianalytic continuation mechanism, provided by the
RGE--evolution of the coupling parameter $a$ itself.
In the one--loop limit, this would
amount to the $[1/1]$ PA--type quasianalytic continuation
mechanism for $a$ itself, which may explain why
especially the ECH method appears to do well even in
the deep nonperturbative regime (where $S$ has large values).

The possibility to adjust the value of the
${\rm N}^3 {\rm LO}$ coefficient $r_3$ of (\ref{TPSBj})
by the ${\rm IR}_1$ pole requirement 
$y_{\rm pole}$ ($\equiv\!2 {\beta}_0 z$) $=\!2$ in the BjPSR
was suggested by the authors of Ref.~\cite{EGKS}.
They chose $r_3$ (at RScl $Q_0^2\!=\!Q^2_{\rm ph}$
and in ${\overline {\rm MS}}$ RSch)
approximately so that the PA $[2/1]$ of the simple Borel
transform of that TPS gave $y_{\rm pole}\!\approx\!2$.
In fact, they chose $r_3\!=\!130.$, which would correspond
to their $y_{\rm pole}\!\approx\!2.10$,
and then resummed the obtained ${\rm N}^3 {\rm LO}$
TPS for $S(Q^2_{\rm ph}\!=\!3 {\rm GeV}^2)$ by
the $[2/2]$ dPA.
However, as we argued in the paragraph following 
Eq.~(\ref{r3pr}), a procedure involving the
simple (RScl-- and RSch--dependent) Borel transform
leads in general to resummed predictions which
can have significant dependence on the RScl and RSch used
in the original TPS (including $c_3^{(0)}$--dependence).
Their approach (with $r_3\!=\!130.$ and $[2/2]$ dPA)
would result in
${\alpha}^{{\overline {\rm MS}}}_s(Q_{\rm ph}^2)\!=\!0.2934^
{+0.0276}_{-0.0370}$ for case I, 
and $0.2891^{+0.0549}_{-0.1058}$ for case II; and
${\alpha}^{{\overline {\rm MS}}}_s(M_Z^2)\!=\!0.1202^
{+0.0040}_{-0.0062}$ for case I and
$0.1140^{+0.0068}_{-0.0201}$ for case II.
Comparing with results in Table \ref{tabl2}, 
we see that these predictions
are again very close to the predictions of $[2/2]_{S^2}^{1/2}$, 
the latter being based solely on the NNLO TPS (\ref{TPSBj}).

Recently, in the context of the Borel--Pad\'e method of resummation
(not used here), the knowledge of the location of renormalon poles
was used in Ref.~\cite{JWS}, in two physical examples,
to fix the denominator structure of the PA's of the
Borel transform.

\section{BjPSR: using Pad\'e--resummed $\beta$--functions}

Since nonperturbative physics appears to be of
high relevance for the high--precision
predictions in the case of the considered observable,
one may go still one step further. Until now,
we used for the $\beta$--functions appearing in
the integrated RGE (\ref{Stforaj}) [cf. also (\ref{RGE})]
simply their TPS to the known order:
\begin{equation}
{\rm TPS}_{\beta}(x) = - {\beta}_0 x^2 (1 + c_1 x 
+ {\bar c}_2 x^2 + {\bar c}_3 x^3) \ ,
\label{TPSb}
\end{equation}
where $x\!\equiv\!{\alpha}_s/\pi$, and the bar over symbols
denotes that they are different in different RSch's. However, 
in the nonperturbative region of large $x$, these TPS's
may give wrong numerical results. To address this question,
we may instead construct PA's based on these TPS's.
PA's represent approximate analytic continuations (i.e., 
quasianalytic continuations) for the true ${\beta}(x)$--functions
from the perturbative (small--$x$) into the nonperturbative 
(large--$x$) region. 
A comprehensive source on
mathematical properties of PA's is the book \cite{Baker}.
We have for (\ref{TPSb}) three PA candidates: 
$[2/3]_{\beta}$, $[3/2]_{\beta}$ and 
$[4/1]_{\beta}$. Constructing these PA's on the basis
of the TPS (\ref{TPSb}), and then reexpanding in powers of $x$,
gives us the higher order RSch--parameters $c_j$ ($j\geq 4$)
that were up until now simply set equal to zero.
Only our approximant $\sqrt{{\cal A}_{S^2}}$, 
and the ECH and TPS--PMS approximants
for the NNLO TPS's (\ref{S2TPS}), are sensitive to this change.
Predictions ${\alpha}_s^{{\overline {\rm MS}}}(Q_{\rm ph}^2)$
of Pad\'e resummation approximants for $S(Q^2_{\rm ph})$
in the previous Section, and the TPS evaluations
themselves (NNLO, ${\rm N}^3 {\rm LO}$), are not
affected by this change (they were calculated
in ${\overline {\rm MS}}$ RSch and at RScl 
$Q^2_0\!=\!Q^2_{\rm ph}$, $n_f\!=\!3$).

For the approximant $\sqrt{{\cal A}_{S^2}}$ the relevant
RSch's are those of $a_1$ (RSch1) and $a_2$ (RSch2), i.e., 
those with the RSch--parameters $(c_2^{(1)}, c_3, \ldots)$
and $(c_2^{(2)}, c_3, \ldots)$, where the dots stand for
$c_k^{(1)}$ and $c_k^{(2)}$ ($k \geq 4$) as determined
by our choice of PA for the RSch1 and RSch2 $\beta$--functions,
respectively. Analogously, for the ECH and TPS--PMS approximants,
the RSch's are $({\rho}_2,c_3,\ldots)$ and 
$(3{\rho}_2/2,c_3,\ldots)$, where the dots 
stand for those RSch--parameters determined by
our choice of the PA for the ECH and TPS--PMS $\beta$--functions. 
So, each of the three choices of the PA defines,
by the afore--mentioned mechanism of quasianalytic continuation
into the nonperturbative sector,
the unique schemes RSch1, ECH RSch, TPS--PMS RSch,
and ${\overline {\rm MS}}$. 

For RSch2, we have to keep in mind one detail:
In order to avoid presumably unnecessary complications, 
the PMS conditions (\ref{PMSeq1})--(\ref{PMSeq2})
were written and used for the choice $c_4^{(2)}\!=\!c_4^{(1)}$
($\delta c_4\!=\!0$), so that the solutions
(\ref{ymres})--(\ref{c2res}) for $Q_1$, $Q_2$, $c_2^{(1)}$
and $c_2^{(2)}$ were independent of 
$c_3$($=\!c_3^{(1)}\!=\!c_3^{(2)}$) and of all the
other $c_k^{(j)}$ ($k \geq 4$; $j\!=\!1,2$). Therefore, 
once we choose a specific $[M/N]_{\beta}$ of the RSch1, the
predicted $c_4$ must be reproduced also by the
$[M^{\prime}/N^{\prime}]_{\beta}$ of the RSch2.
This means that the order of the latter PA is
by one unit higher than that of the former:
$M^{\prime}\!+\!N^{\prime}\!=\!M\!+\!N\!+\!1$.
Since the PA choices for the RSch1 $\beta$--function
are [2/3], [3/2] and [4/1], those for the RSch2 $\beta$--function are: 
[2/4], [3/3], [4/2], [5/1].
As to the numerics, the situation does not change much
when different choices of $[M^{\prime}/N^{\prime}]_{\beta}$ 
or even TPS for the RSch2 are taken (with $c_4^{(2)}\!=\!c_4^{(1)}$,
and always the same fixed value of $c_3$). 
This is so because, in the strong--coupling regimes
$S \geq 0.155$, $a_1$ is by a factor of $1.66$ or more 
larger than $a_2$.
Concerning the choice of ${\rm PA}_{\beta}$ of 
${\overline {\rm MS}}$ RSch, this choice does not influence
the predictions of $c_3$ at all, and influences
only little the subsequent predictions for 
${\alpha}_s^{{\overline {\rm MS}}}(Q_{\rm ph}^2)$.
The latter is true mainly because of the hierarchy:
$a_0 < a_2 < a_1$ ($Q_0 > Q_2 > Q_1$: 
$Q_1\!\approx\!0.343 Q_{\rm ph}$,
$Q_2\!\approx\!0.672 Q_{\rm ph}$,
$Q_0\!=\!Q_{\rm ph}\!\approx\!1.73$ or $2.24$ GeV).

For the various ${\rm PA}_{\beta}$ choices of RSch1, RSch2,
ECH RSch and TPS--PMS RSch, we can just redo the
entire calculation of the invariant Borel transforms
$\overline{B}_S$ of (\ref{BT3}) and of their PA's,
and find predictions for $c_3$ that give us
the correct ${\rm IR}_1$ pole 
$y_{\rm pole}\!=\!2$. It turns out that the most
stable $c_3$--predictions in our approximant
$\sqrt{{\cal A}_{S^2}}$ are those with
$[2/3]_{\beta 1}$ for RSch1 (${\beta}1$) and 
$[2/4]_{\beta 2}$ for RSch2 (${\beta}2$).
The choice $[2/3]_{\beta 1}$ and $[5/1]_{\beta 2}$
gives virtually the same and almost as stable
predictions for $c_3$.
For the ECH and TPS--PMS approximants, all three
choices $[2/3]_{\beta}$, $[3/2]_{\beta}$, and
$[4/1]_{\beta}$ give comparably stable and mutually quite
similar $c_3$--predictions, but the choice $[3/2]_{\beta}$
seems to be slightly more stable than the other two. 
The results, for the mentioned optimal
choices of ${\rm PA}_{\beta}$'s for the three
approximants, are given in Table \ref{tabl3}, in
complete analogy with Table \ref{tabl1}.
In some cases there are also other solutions for $c_3$, 
not included in the Table, which differ significantly from 
those given in the Table. 
\begin{table}[ht]
\par
\begin{center}
\begin{tabular}{l c | c | c }
${\rm PA}_{\overline{B}}$  & 
$c_3$ for $\sqrt{\cal A}$: $[2/3]_{{\beta}1}$, $[2/4]_{{\beta}2}$ & 
$c_3$ for ECH: $[3/2]_{\beta}$ & $c_3$ for TPS--PMS: $[3/2]_{\beta}$ \\
\hline \hline
$[2/1]$ & 21.7 & 35.1 & 35.1 \\
\hline
$[3/1]$ & 15.7 & 22.9 & 21.5 \\
\hline
$[4/1]$ & 15.8 & 20.8 & 18.7 \\ 
\hline
$[5/1]$ & 16.9 & 19.6 & 17.3 \\
\hline\hline
$[1/2]$ & 12.8 & 17.3 & 17.3 \\
\hline
$[2/2]$ & 14.9 & 20.4 & 19.4 \\ 
\hline
$[3/2]$ & 15.8 & $20.7\!\pm\!2.8 {\rm i}$ & $17.3\!\pm\!3.6 {\rm i}$ \\
\hline
$[4/2]$ & 15.7 & $20.4\!\pm\!1.8 {\rm i}$ & $17.0\!\pm\!2.6 {\rm i}$ \\
\hline\hline
$[1/3]$ & 15.0 & 20.6 & 19.5 \\
\hline
$[2/3]$ & $15.1\!\pm\!1.2 {\rm i}$ & 19.3 & 18.5 \\ 
\hline
$[3/3]$ & $14.0\!\pm\!1.7 {\rm i}$ & $20.2\!\pm\!2.0 {\rm i}$ & 
$16.9\!\pm\!2.7 {\rm i}$ \\
\hline\hline
average & $\approx 15.5$ & $\approx 20.$ & $\approx 19.$
\end{tabular}
\end{center}
\caption {\footnotesize  As in Table \ref{tabl1}, but the
$\beta$--functions in the approximants are taken as:
$[2/3]_{\beta}$ (RSch1), 
$[2/4]_{\beta}$ (RSch2; $c_4^{(2)}\!=\!c_4^{(1)}$);
$[3/2]_{\beta}$ (ECH RSch, and TPS--PMS RSch).}
\label{tabl3}
\end{table}
We will adopt the approximate predictions as suggested
by ${\rm PA}_{\overline{B}}$'s of intermediate orders
([3/1], [4/1], [2/2], [3/2], [1/3], [2/3]):
$c_3 \approx 15.5$ for $\sqrt{{\cal A}_{S^2}}$;
$c_3 \approx 20$ for the ECH;
$c_3 \approx 19$ for the TPS--PMS.
The actual values of $c_3$ must be exactly real.

We recall that the results of the previous two Sections, 
including those of Table \ref{tabl1},
were for the simple choice of ${\rm TPS}_{\beta}$ (\ref{TPSb})
for the corresponding RSch's (``truncated RSch's,'' 
with $c_k\!=\!0$ for $k\geq4$). Comparing those results
with the results of Table \ref{tabl3}, we see
that the latter are somewhat higher and significantly more stable
under the change of the choice of ${\rm PA}_{\overline{B}}$.
This latter fact can be regarded as a numerical indication 
that it makes sense to use certain PA resummations for the
pertaining $\beta$--functions of approximants
when the considered observable (in this case BjPSR)
contains nonperturbative effects.

When the order of ${\rm PA}_{\overline{B}}$ is increased,
the trend of the predictions is similar as in
Table \ref{tabl1}: The predictions $c_3$ tend to stabilize
at intermediate orders of the ${\rm PA}_{\overline{B}}$'s.
The lowest order ${\rm PA}_{\overline{B}}$'s ($[1/2]$, and
above all $[2/1]$) give unreliable predictions for $c_3$,
apparently because of a too simple structure of 
these PA's. The highest order ${\rm PA}_{\overline{B}}$'s
($[5/1]$, $[4/2]$, $[3/3]$) also sometimes
give unreliable predictions,
apparently because of their ``overkill'' capacity --
these ${\rm PA}_{\overline{B}}$'s depend on many terms
in the power expansion of the approximant 
(up to $\sim$${\tilde a}^7$), while the original TPS 
(\ref{TPSBj}) on which the approximant is based
is given only up to $\sim$$a_0^3$ ($\sim$${\tilde a}^3$).
Therefore, it seems plausible that the best and most stable
predictions are given by ${\rm PA}_{\overline{B}}$'s of
intermediate orders ($[3/1]$, $[4/1]$, $[2/2]$, $[3/2]$,
$[1/3]$, $[2/3]$).

With these choices for the values of $c_3$ and for the
pertaining $\beta$--functions, we could now go on to
calculating predictions 
of the three approximants for 
${\alpha}_s^{{\overline {\rm MS}}}$. 
Since the choice of ${\rm PA}_{\beta}$ for
${\overline {\rm MS}}$ RSch will not matter
much numerically, as we argued above, we could just
choose blindly a ${\rm PA}_{\beta}$ or even the TPS for it.
But at this point, we want to point out an additional
argument for the made ${\rm PA}_{\beta}$ choices of
RSch1/RSch2, ECH RSch and TPS--PMS RSch.
This argument will, in addition, lead us to a specific
choice of ${\rm PA}_{\beta}$ for ${\overline {\rm MS}}$ RSch.

\begin{figure}[ht]
 \centering\epsfig{file=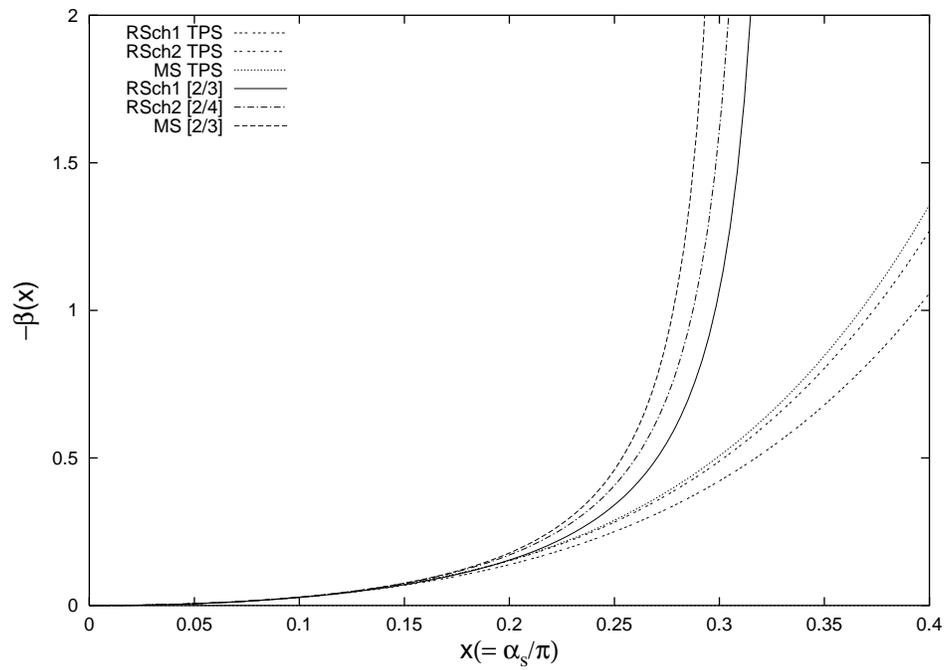}
\vspace{0.3cm}
\caption{\footnotesize
TPS $\beta$--functions for RSch1 and RSch2 ($c_3\!=\!15.5$),
and ${\overline {\rm MS}}$ ($n_f\!=\!3$), and their corresponding
PA's $[2/3]$, $[2/4]$ ($c_4^{(2)}\!=\!c_4^{(1)}$),
and $[2/3]$, respectively.}
\label{fig2Bjf}
\end{figure}

In this context, we recall first that quasianalytic 
continuation, e.g. via PA's,
of the TPS of a $\beta$--function into the 
large--$x$ (nonperturbative) region leads in general
to a pole of such ${\rm PA}_{\beta}(x)$ at some
positive $x$. The authors of Ref.~\cite{Chishtieetal}
pointed out that these poles ``suggest the occurrence
of dynamics in which both a strong and an 
asymptotically--free phase share a common infrared attractor.''
Now, if there is such a common point 
$x_{\rm pole}\!\equiv\!{\alpha}_s^{\rm pole}/\pi$
where the two phases meet, it is reasonable to expect
that its numerical value does not vary wildly
when we change RSch -- provided that the RSch's
in question are themselves physically motivated
(physically reasonable) in the 
nonperturbative regime.\footnote{
In the perturbative regime, all RSch's are formally
equivalent.} 
Such physically motivated
RSch's should include those connected
in some significant way with the calculation of
the considered observable and of the predicted
coupling parameters. In the case of our approximant
$\sqrt{{\cal A}_{S^2}}$, these are RSch1 and RSch2,
and in addition ${\overline {\rm MS}}$ when we want
to extract ${\alpha}_s^{{\overline {\rm MS}}}(Q_{\rm ph}^2)$
from the approximant. In Fig.~\ref{fig2Bjf} we present the TPS's
of RSch1, RSch2 and ${\overline {\rm MS}}$
$\beta$--functions, as well as the previously
chosen $[2/3]_{\beta 1}$ of RSch1 and
$[2/4]_{\beta 2}$ of RSch2 (cf.~Table \ref{tabl3}; $c_3=15.5$),
and we include also $[2/3]_{\beta}$ of ${\overline {\rm MS}}$
RSch. The Figure shows that all these 
${\rm PA}$ $\beta$--functions have about the same
$x_{\rm pole}$ ($x_{\rm pole}\!=\!0.334, 0.325, 0.311$,
respectively). The mutual proximity of $x_{\rm pole}$'s of
RSch1 and RSch2 ${\rm PA}_{\beta}$'s is now
yet another indication that these ${\rm PA}_{\beta}$'s,
chosen previously on the basis of the stability of
$c_3$--predictions, are the reasonable ones. Further,
$[2/3]_{\beta}$ appears to be the reasonable
choice for ${\overline {\rm MS}}$ RSch.
The choices $[3/2]_{\beta}$ and $[4/1]_{\beta}$
for  ${\overline {\rm MS}}$ RSch give
$x_{\rm pole}\!=\!0.119, 0.213$, respectively,
which is further away from the $x_{\rm pole}$
of RSch1 and RSch2. We could choose, in principle, 
for RSch1 and RSch2 other ${\rm PA}_{\beta}$'s.
We recall that for RSch1 we can have: $[2/3]_{\beta 1}$, 
$[3/2]_{\beta 1}$, $[4/1]_{\beta 1}$; for RSch2: 
$[2/4]_{\beta 2}$, $[3/3]_{\beta 2}$, 
$[4/2]_{\beta 2}$, $[5/1]_{\beta 2}$.
However, when taking
$[3/2]_{\beta 1}$ or $[4/1]_{\beta 1}$, we always end up 
either with a situation when the two positive $x_{\rm pole}$ 
values of ${\beta} 1$ and ${\beta 2}$ are far apart,
or both are unphysically small, or one positive $x_{\rm pole}$ 
doesn't exist, or there are virtually no predictions for $c_3$
(not even unstable ones), or $x_{\rm pole}$ values are very unstable
under the change of $c_3$ in the interesting region
$c_3\!\approx\!12$--$16$.
Concerning the latter point -- when taking $[3/2]_{\beta 1}$,
and for RSch2 $[3/3]_{\beta 2}$ or $[4/2]_{\beta 2}$,
the location of $x_{\rm pole}$ of the latter ${\rm PA}_{\beta}$'s
changes drastically when $c_3$ is varied around the 
interesting values of $12$--$16$, thus
signalling instability of these ${\rm PA}_{\beta}$'s. 
The choice $[2/3]_{\beta 1}$
and $[5/1]_{\beta 2}$, which gave
very similar and almost as stable results for $c_3$
as the most preferred choice $[2/3]_{\beta 1}$ 
and $[2/4]_{\beta 2}$,
gives the corresponding poles again close to each
other: $x_{\rm pole}\!=\!0.334, 0.291$, respectively.
So, the ${\rm PA}_{\beta}$
choices $[2/3]_{\beta 1}$ and 
$[2/4]_{\beta 2}$ (or $[5/1]_{\beta 2}$)
for our approximant give us the most stable
$c_3$--predictions {\em and} are the only ones
giving mutually similar (and reasonable) values of $x_{\rm pole}$
of RSch1 and RSch2.

It is also encouraging that the choices
$[3/2]_{\beta}$ for the ECH and TPS--PMS RSch's
give us $x_{\rm pole}$ values comparable
to the ones previously mentioned:
$x_{\rm pole}\!=\!0.263$ for ECH with $c_3\!=\!20$;
$x_{\rm pole}\!=\!0.327$ for TPS--PMS with $c_3\!=\!19$.
Even other choices of ${\rm PA}_{\beta}$
for the ECH and TPS--PMS RSch's ($[2/3]_{\beta}$, $[4/1]_{\beta}$),
which also gave rather stable and similar $c_3$--predictions,
give us $x_{\rm pole} \approx 0.27$--$0.41$.
Hence, also in this case we see correlation
between the stability of the $c_3$--predictions
on the one hand and $x_{\rm pole} \approx 0.3$--$0.4$
on the other hand.

The authors of Refs.~\cite{c3pr,c4pr} estimated the
5--loop coefficient $c_4^{{\overline {\rm MS}}}$
of the ${\overline {\rm MS}}$ $\beta$--function, by
applying their method of Asymptotic Pad\'e Approximation
(APAP, \cite{c3pr}) and its improvement using estimators
over negative numbers of flavors (WAPAP, \cite{c4pr}).
Their predicted values by two variants of the latter method,
when including the four--loop quartic Casimir contributions, 
are $c_4^{{\overline {\rm MS}}}\!=\!123.7, 115.3$
(cf. Tables III and IV in Ref.~\cite{c4pr}, respectively; 
$n_f\!=\!3$). On the other hand, the simple PA's 
$[2/3]$, $[3/2]$, $[4/1]$ for ${\overline {\rm MS}}$ 
$\beta$--function predict $c_4^{{\overline {\rm MS}}}\!=\!62.2, 
149.8, 98.5$,  and $x_{\rm pole}\!=\!0.311, 0.119, 0.213$, 
respectively. If we assume that the actual
value of $c_4^{{\overline {\rm MS}}}$ is close to the one
predicted by \cite{c4pr}, and if we were led just by the
requirement that the PA should reproduce well this value,
then $[4/1]$ would be the preferred choice. However, the
authors of \cite{c4pr} indicated that their predicted
value of $c_4$ may be changed significantly if new
Casimir terms, appearing for the first time at the 5--loop order, 
are large. Our choice $[2/3]$ for ${\overline {\rm MS}}$ 
$\beta$--function was motivated by the value of $x_{\rm pole}\!=\!0.311$
lying close to $x_{\rm pole}$ of the $\beta$--functions
appearing in the discussed approximants for the BjPSR.
Further, the precise choice of the PA for
${\overline {\rm MS}}$ $\beta$--function practically does not
influence the numerical results of our analysis, because
$a_0\!\equiv\!a(\ln Q_{\rm ph}^2; c_2^{{\overline {\rm MS}}}, \ldots)$
is significantly smaller than the coupling parameters
$a_j\!\equiv\!a(\ln Q_j^2; c_2^{(j)}, c_3, c_4, c_5^{(j)},\ldots)$ 
($j\!=\!1,2$) appearing in our approximant,
and the parameters $a_{\rm ECH}$ and $a_{\rm PMS}$
appearing in the ECH and the TPS--PMS approximants.

To summarize: 
\begin{itemize}
\item
the best choice in calculating
${\alpha}_s^{{\overline {\rm MS}}}$ from
our approximant $\sqrt{{\cal A}_{S^2}}$ is: 
$c_3\!\approx\!15.5$; the ${\rm PA}_{\beta}$ choice 
$[2/3]_{\beta}$ for RSch1, $[2/4]_{\beta}$
choice for RSch2 ($c_4^{(2)}\!=\!c_4^{(1)}$);
and $[2/3]_{\beta}$ for ${\overline {\rm MS}}$ RSch;
\item
the best choice in calculating 
${\alpha}_s^{{\overline {\rm MS}}}$ from
the ECH and TPS--PMS approximants is:
$c_3\!\approx\!20$ and $19$, respectively;
the ${\rm PA}_{\beta}$ choice $[3/2]_{\beta}$
for ECH RSch and TPS--PMS RSch; and
$[2/3]_{\beta}$ for ${\overline {\rm MS}}$ RSch;
\item
our, the ECH and the TPS--PMS approximants
are completely independent of the original choice of the
RScl and RSch, because the $c_3$ parameter is determined
by using the RScl-- and RSch--invariant Borel transform
${\overline B}(z)$ of Sec.~III.

\end{itemize}

In practice, this means that for our approximant
$\sqrt{{\cal A}_{S^2}}$ the two coupling parameters 
$a_j\!\equiv\!a(\ln Q_j^2; c_2^{(j)}, c_3, c_4, c_5^{(j)},\ldots)$ 
($j\!=\!1,2$) are now related with the
coupling parameter 
$a_0\!\equiv\!a(\ln Q_0^2; c_2^{{\overline {\rm MS}}},
c_3^{{\overline {\rm MS}}}, c_4^{{\overline {\rm MS}}}, \ldots)$ 
via the following
(PA--)version of the subtracted Stevenson equation (\ref{Stforaj})
[cf. also (\ref{Stevenson})--(\ref{Stsubtr})]:
\begin{eqnarray}
\lefteqn{
\beta_0 \ln \left( \frac{Q_j^2}{Q_0^2} \right) = 
\frac{1}{a_j} + c_1 \ln \left( \frac{ c_1 a_j}{1\!+\!c_1 a_j} \right)
+ \int_0^{a_j}\!dx 
\frac{ \left\{ {\rm PA}_{\beta j}(x) + 
\beta_0 x^2 ( 1\!+\!c_1 x) \right\} }
{ x^2 ( 1\!+\!c_1 x) {\rm PA}_{\beta j}(x) }
}
\nonumber\\
&& - \frac{1}{a_0} - 
c_1 \ln \left( \frac{ c_1 a_0}{1\!+\!c_1 a_0} \right)
- \int_0^{a_0}\!dx
\frac{ \left\{ [2/3]_{{\rm MS} \beta}(x) + 
\beta_0 x^2 ( 1\!+\!c_1 x) \right\} }
{ x^2 ( 1\!+\!c_1 x) [2/3]_{{\rm MS} \beta}(x) } \ ,
\label{Stsubtrgen}
\end{eqnarray}
where ${\rm PA}_{\beta j}$ stands for the mentioned
$[2/3]_{\beta}$ of RSch1 (when $j\!=\!1$)
and $[2/4]_{\beta}$ of RSch2 (when $j\!=\!2$), with $c_3=15.5$.
We recall that the scales $Q_j^2$ and the parameters
$c_2^{(j)}$ ($j\!=\!1,2$) of the approximant, 
which are RScl-- and RSch--invariant
and calculated in Sections II and III 
[cf.~(\ref{ymres})--(\ref{c2res})], are independent
of the parameter $c_3$ and of any higher order $\beta$--parameter
$c_k^{(j)}$ ($k \geq 4$; $j\!=\!1,2$) appearing in
$a_j\!\equiv\!a(\ln Q_j^2; c_2^{(j)}, c_3, c_4, c_5^{(j)},\ldots)$.
For the ECH and TPS--PMS the calculation is performed
in an analogous way.

The results of these calculations, i.e., the predicted values of
${\alpha}^{{\overline {\rm MS}}}_s(Q_{\rm ph}^2)$
and ${\alpha}^{{\overline {\rm MS}}}_s(M_Z^2)$,
are given in Table \ref{tabl4} for 
the approximants $\sqrt{{\cal A}_{S^2}}$, ECH and TPS-PMS.
\begin{table}[ht]
\par
\begin{center}
\begin{tabular}{l| c c |c c}
approximant (with ${\rm PA}_{\beta}$'s) & 
$\alpha_s( 5 \ {\rm GeV}^2)$: (I) & 
$\alpha_s( 3 \ {\rm GeV}^2)$: (II) & 
$\alpha_s(M_Z^2)$: (I) & $\alpha_s(M_Z^2)$: (II) \\
\hline \hline
$\sqrt{ {\cal A}_{S^2}^{[2/2]} }$ ($c_3\!=\!15.5$) & 
$0.2838^{+0.0182}_{-0.0311}$ &
$0.2805^{+0.0297}_{-0.0977}$ &
$0.1187^{+0.0028}_{-0.0054}$ &
$0.1127^{+0.0041}_{-0.0189}$ \\
\hline
ECH ($c_3\!=\!20.$) &  
$0.2856^{+0.0195}_{-0.0321}$ &
$0.2822^{+0.0325}_{-0.0993}$ &
$0.1190^{+0.0030}_{-0.0056}$ &
$0.1130^{+0.0044}_{-0.0192}$ \\
\hline
TPS--PMS ($c_3\!=\!19.$) & 
$0.2867^{+0.0202}_{-0.0328}$ &
$0.2831^{+ {\rm ?}}_{-0.1001}$ &
$0.1192^{+0.0030}_{-0.0057}$ &
$0.1131^{+ {\rm ?}}_{-0.0192}$ \\ 
\end{tabular}
\end{center}
\caption {\footnotesize Predictions for 
${\alpha}_s^{{\overline {\rm MS}}}$
for our, ECH, and TPS--PMS approximants, when the 
PA--resummed $\beta$--functions
in the approximants are taken as in Table \ref{tabl3}.
Predictions for the case
I (\ref{BjdatI}) and II (\ref{BjdatII}) are given in parallel.}
\label{tabl4}
\end{table}
The predictions are now a little, but still significantly,
lower than those of the corresponding approximants
in Table \ref{tabl2} where all the $\beta$--functions
were taken in the TPS form (\ref{TPSb}) and with 
$c_3\!=\!12.5, 17, 16$, respectively.
The evolution from 
${\alpha}^{{\overline {\rm MS}}}_s(Q_{\rm ph}^2)$
to ${\alpha}^{{\overline {\rm MS}}}_s(M_Z^2)$
was performed as in the previous Section,
i.e., with the four--loop RGE (i.e., TPS
$\beta$--function of ${\overline {\rm MS}}$)
and the corresponding three--loop flavor threshold
matching conditions. If we replace the 
TPS $\beta$--function of ${\overline {\rm MS}}$
by its PA $[2/3]_{\beta}$ in the RGE for the evolution
${\alpha}^{{\overline {\rm MS}}}_s(Q_{\rm ph}^2) \mapsto
{\alpha}^{{\overline {\rm MS}}}_s(M_Z^2)$,
the results for ${\alpha}^{{\overline {\rm MS}}}_s(M_Z^2)$
decrease insignificantly (by less than $0.04 \%$)
and the numbers in Table \ref{tabl4} do not change. 

\begin{figure}[ht]
\centering\epsfig{file=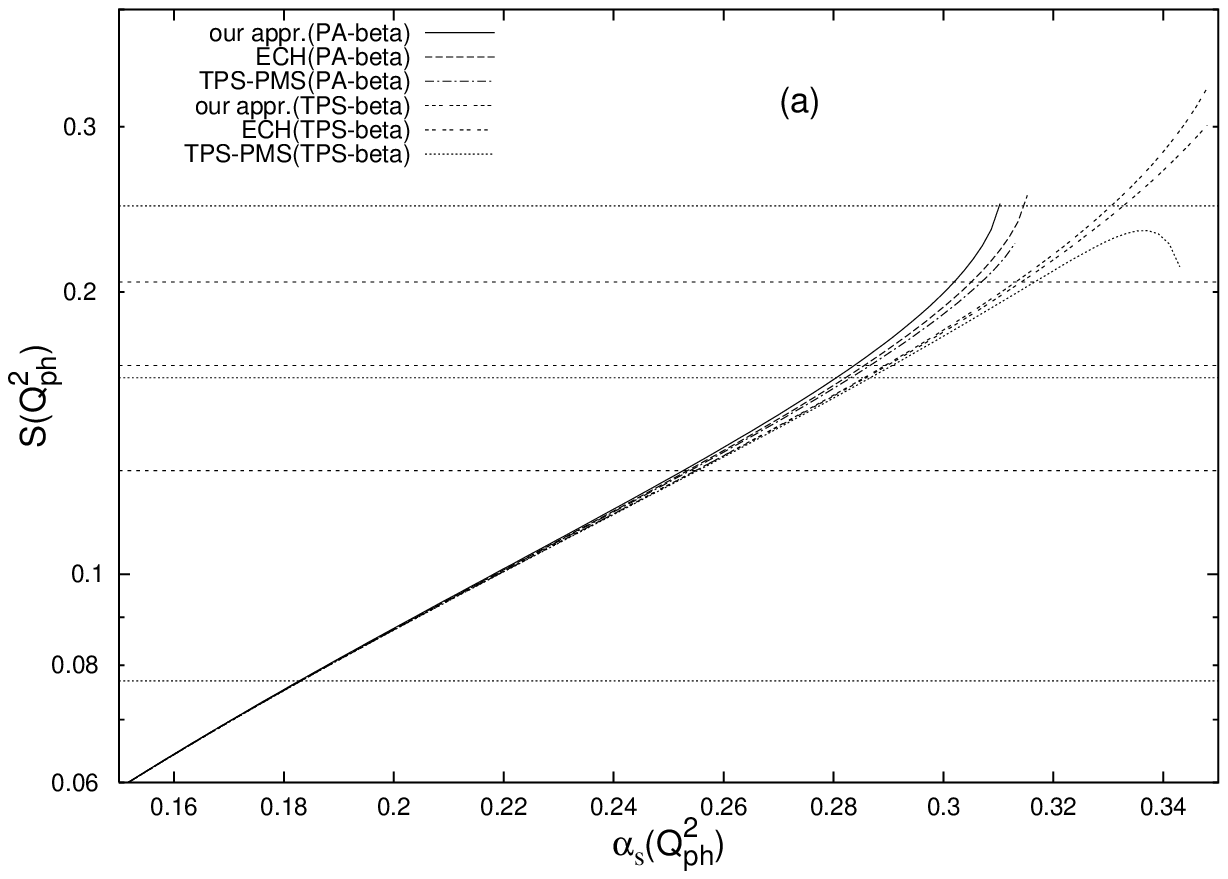,height=9.5cm}
\centering\epsfig{file=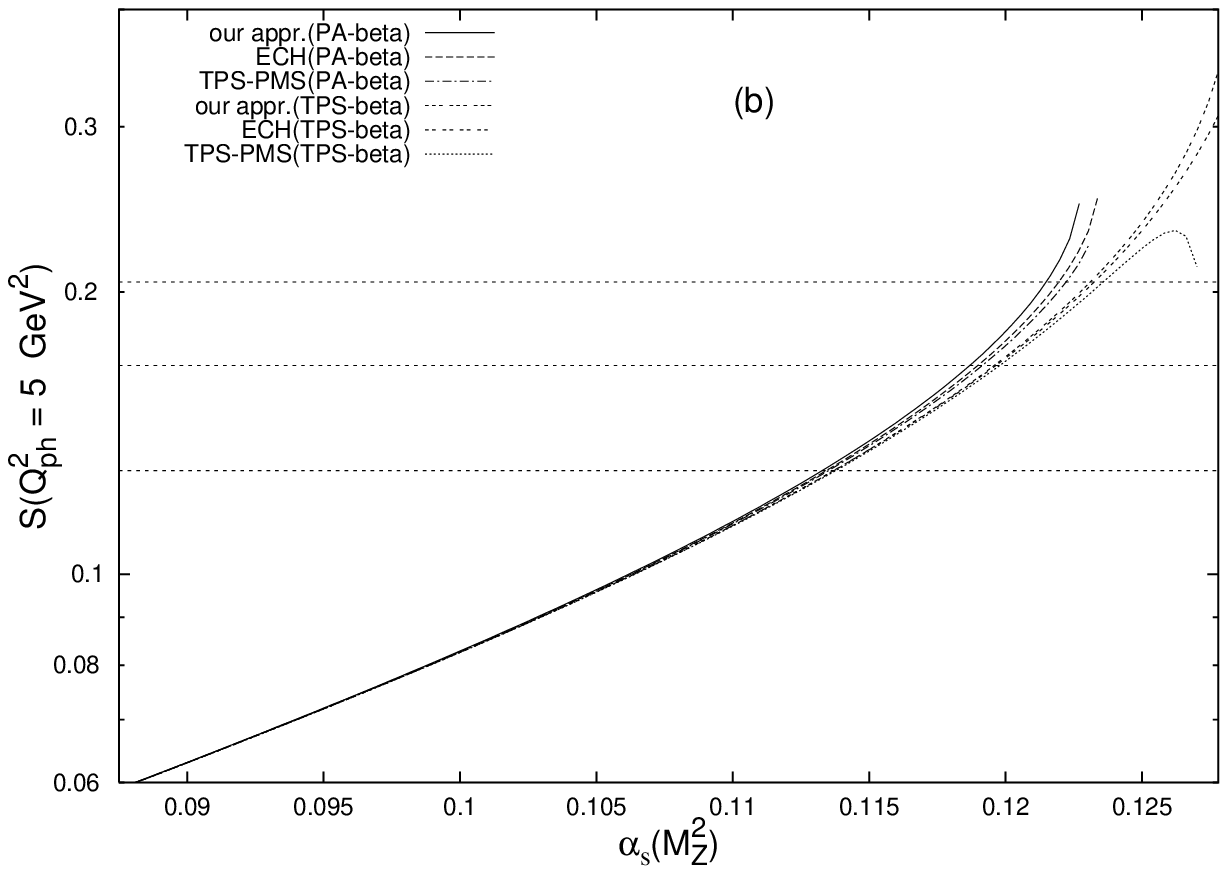,height=9.5cm}
\vspace{0.3cm}
\caption{\footnotesize Predictions of our approximant 
(with $c_3\!=\!15.5$),
ECH (with $c_3\!=\!20.$), 
and TPS--PMS (with $c_3\!=\!19.$):
(a) for $S(Q_{\rm ph}^2)$ as functions of 
${\alpha}^{{\overline {\rm MS}}}_s(Q_{\rm ph}^2)$ when $n_f\!=\!3$;
(b) for $S(Q_{\rm ph}^2\!=\!5{\rm GeV}^2)$ as functions of 
${\alpha}^{{\overline {\rm MS}}}_s(M_Z^2)$.
The PA choices of the RGE $\beta$--functions were made
as explained in the text. For comparison, we include also
the corresponding predictions from Figs.~\ref{fig1Bjf}
when the TPS's (\ref{TPSb}) are used for the $\beta$--functions.
The values of the $c_3$--parameter 
have been adjusted in all cases to ensure
the correct location of the leading IR renormalon pole.
The experimental bounds are denoted as in 
Figs.~\ref{fig1Bjf}.}
\label{fig3Bjf}
\end{figure}

In Fig.~\ref{fig3Bjf}(a) we present
predictions $S(Q^2_{\rm ph})$ as functions of 
${\alpha}^{{\overline {\rm MS}}}_s(Q_{\rm ph}^2)$
($n_f\!=\!3$, e.g. $Q_{\rm ph}^2\!=\!3$ or $5 \ {\rm GeV}^2$), 
and in Fig.~\ref{fig3Bjf}(b) the predictions for
$S(5 {\rm GeV}^2)$ as functions of
${\alpha}^{{\overline {\rm MS}}}_s(M_Z^2)$,
for the three approximants with the
afore--mentioned PA choices for the $\beta$--functions. 
For comparison,
we include in the Figures also predictions of these
three approximants when all the $\beta$--functions have the
TPS form (\ref{TPSb}) and the correspondingly smaller $c_3$'s
(the latter curves are contained also in Figs.~\ref{fig1Bjf}). 
Predictions of the PA resummation approximants (for $S$) are not
included, since these methods are insensitive to
the mentioned PA--quasianalytic continuation
of the $\beta$--functions and the results
remain for them the same as in Figs.~\ref{fig1Bjf}
and Table \ref{tabl2}.
We presented in Figs.~\ref{fig3Bjf}
the curves for the case of approximants with the
mentioned PA $\beta$--functions
only so far as the method works. More specifically, when the
integration interval in the first integral
of (\ref{Stsubtrgen}) 
starts including values $x$ larger than 
those at which the absolute value of the
${\rm PA}_{\beta}$ exceeds the value $2$,
we stop the calculation of the approximant
since the latter would otherwise probe
values too near the pole of ${\rm PA}_{\beta}$
(i.e., too near the common point of the 
asymptotically--free and the strong phase)
and would thus be unreliable.

The considered BjPSR observable $S(Q^2_{\rm ph})$
has a higher--twist (h.t.\/) contribution,
estimated from QCD sum rule \cite{htlit}\footnote{
Deficiencies of the QCD sum rule calculations
were pointed out in \cite{Ji}.}
\begin{equation}
S^{\rm (h.t.)}(Q^2_{\rm ph}) \approx 
\frac{ (0.09 \pm 0.045) \ {\rm GeV}^2 }{Q^2_{\rm ph}} \ ,
\label{ht}
\end{equation}
which should be added to the perturbation series for $S$. 
If adding this term 
in the numerical analysis, the predicted central
values of ${\alpha}^{{\overline {\rm MS}}}_s(M_Z^2)$
given in Table \ref{tabl2}
decrease significantly.
For example, the NNLO TPS central value predictions 
${\alpha}^{{\overline {\rm MS}}}_s(M_Z^2)\!=\!0.1252$ (case I)
and $0.1183$ (case II)
then decrease to $0.1200$--$0.1236$ (case I)
and $0.1091$--$0.1157$ (case II), where the
lower and upper values in each case correspond to the largest
and the smallest value choice in (\ref{ht}).
This indicates numerically that our approximant 
($c_3\!=15.5$, Table \ref{tabl4}),
which gives the central values 
${\alpha}^{{\overline {\rm MS}}}_s(M_Z^2)\!=\!0.1187$ (case I)
and $0.1127$ (case II), already contains at 
least part of the nonperturbative effects
from the leading higher--twist operator
($\sim$$1/Q^2_{\rm ph}$). The same is true for the ECH
($c_3\!=\!20.$) and TPS--PMS($c_3\!=\!19.$).  
In order to understand this numerical
indication, we recall that the information on the
location of the leading IR renormalon (${\rm IR}_1$)
pole of the considered observable has
already been incorporated in these approximants,
via the afore--mentioned fixing of the
value of $c_3$--parameter. And the
so called ambiguity of the leading IR
renormalon is of the same
form $\sim$$1/Q^2_{\rm ph}$ as the higher--twist
term (\ref{ht}), and even the estimated
coefficients are of the same order of
magnitude \cite{Braun} (cf.~also Ref.~\cite{EGKS} on this point).
Our approximant, the ECH and the TPS--PMS,
via the discussed $c_3$--fixing, implicitly provide 
approximant--specific prescriptions of how to integrate
in the Borel integral over the ${\rm IR}_1$ pole,
thus eliminating the (leading) renormalon ambiguity.

\section{Discussion of the numerical results}

The main reason to apply
our approach (and PA approaches) to the BjPSR was to
investigate efficiencies of various methods and the
influence of the nonperturbative sector. Another
reason was that the BjPSR is a Euclidean observable
($q^2_{\rm ph}\!=\!-Q^2_{\rm ph} < 0$), and for
such observables various resummation methods are believed to
work well since no real particle thresholds are involved
in the observable \cite{KS,CKS}.

The main prediction of our approximant
$\sqrt{{\cal A}_{S^2}}$
can be read off from Table \ref{tabl4},
for two cases (\ref{BjdatI}) and (\ref{BjdatII})
of the BjPSR--integral values at 
$Q^2_{\rm ph}\!=\!5$ and $3 \ {\rm GeV}^2$, respectively,
extracted from experiments 
\begin{equation}
{\alpha}^{{\overline {\rm MS}}}_s(M_Z^2) = 
0.1187^{+0.0028}_{-0.0054} \ (I); \quad
0.1127^{+0.0041}_{-0.0189} \ (II).
\label{result}
\end{equation}
The ECH and the TPS--PMS give results
similar to these, when $c_3$--parameter
in them is adjusted in the afore--mentioned way
-- see Table \ref{tabl4}.
The diagonal PA (dPA) methods give higher predictions,
and the nondiagonal PA methods even higher -- see Table \ref{tabl2}
and Figs.~\ref{fig1Bjf}.

The result (\ref{result}) for case II, which is based on the
measurements before 1997 and a NLO pQCD
extrapolation for low $x_{\rm Bj}$ \cite{ABFR} (\ref{BjdatII}), 
shows quite large uncertainties, a consequence of
the large uncertainties (\ref{SII}).
[$\Leftrightarrow$ (\ref{BjdatII})]. 
The result (\ref{result}) for case I,
based on the most recent measurements and 
a similar NLO pQCD extrapolation for small $x_{\rm Bj}$,
by the SLAC E155 Collaboration \cite{2000E155} (\ref{BjdatI}),
already shows significantly reduced uncertainties.
This is so to a large degree because of additional
new measurements in the low-$x_{\rm Bj}$ regime.
And most importantly, the central values of 
case I in (\ref{result}) are now significantly higher
than those of (the older) case II. 
We recall that the central values in (\ref{result}) correspond 
to the central values of the BjPSR--integral (\ref{BjdatI})
and (\ref{BjdatII}). We did not attempt to estimate the
theoretical uncertainties originating from the resummation
method itself. However, the combined results of
Table \ref{tabl4} 
${\alpha}^{{\overline {\rm MS}}}_s(M_Z^2)\!=\!0.119^{+0.003}_{-0.006}$
for (new) case I could be regarded as containing 
nonconservatively estimated theoretical uncertainties.

The present world average is
${\alpha}_s^{{\overline {\rm MS}}}(M_Z^2) =
0.1173 \pm 0.0020$ by Ref.~\cite{Hinchliffe},
and $0.1184 \pm 0.0031$ by Ref.~\cite{Bethke}.
Predictions of the simple (NNLO) TPS evaluation
in (new) case I give $0.1252^{+0.0055}_{-0.0078}$ 
(see Table \ref{tabl2}),
the central value and most of the interval lying
significantly above the world average.
On the other hand, the simple (NNLO) TPS evaluation
in (older) case II predicts 
$0.1183^{+0.0095}_{-0.0232}$ (see Table \ref{tabl2}),
the central value agreeing well with the world average,
but the uncertainty interval being much broader.
However, the situation changes drastically
when employing more sophisticated resummation methods.
The values for BjPSR--predicted 
${\alpha}_s^{{\overline {\rm MS}}}(M_Z^2)$
go down the more significantly, the more sophisticated
resummation we perform -- cf.~Table \ref{tabl2}
for the PA--methods, and for TPS--PMS, ECH and
$\sqrt{{\cal A}_{S^2}}$ when the $\beta$--functions have
truncated form, and Table \ref{tabl4} for the last
three methods when the $\beta$--functions are resummed.
{}The predictions of approximants in the latter Table
have ${\alpha}_s^{{\overline {\rm MS}}}(M_Z^2)\approx 
0.119^{+0.003}_{-0.006}$  (case I, new) and 
$0.113^{+0.004}_{-0.019}$ (case II, old).
The predictions of (new) case I now agree well
with the world average $0.1184 \pm 0.0031$ of Ref.~\cite{Bethke},
while those of (older) case II lie almost entirely below the
world average intervals.

Thus, the use of resummation
methods which account for nonperturbative
contributions by the mechanism of quasianalytic
continuation and by incorporation of the information
on the leading IR renormalon pole, predict 
the values of ${\alpha}_s^{{\overline {\rm MS}}}(M_Z^2)$
which agree well with the present world average
if the most recent BjPSR data \cite{2000E155} are used.
This suggests, among other things, that for reliable 
predictions of ${\alpha}^{{\overline {\rm MS}}}_s$
from reasonably well measured low--energy QCD observables,
we have to know the NNLO terms ($\sim\!a^3$), employ
nontrivial resummation methods, and possibly incorporate
some nonperturbative (renormalon) information in the resummation.

Some of the recently performed analyses beyond the NLO,
by other authors, gave predictions:
${\alpha}_s^{\overline {\rm MS}}(M_Z^2) =
0.118 \pm 0.006$ \cite{Kataevetal} 
from the CCFR data for $x_{\rm Bj} F_3$
structure function from ${\nu} N$ DIS (NNLO);
$0.1172 \pm 0.0024$ \cite{SY} from ${\ell} N$ DIS (NNLO);
$0.115 \pm 0.008$ \cite{GLS1} and 
$0.114^{+0.010}_{-0.012}$ \cite{GLS2} from
Gross--Llewellyn--Smith sum rule (NNLO);
$0.1181 \pm 0.0031$ from hadronic $\tau$--decay
(NNLO, combined results, \cite{Bethke});
$0.115 \pm  0.004$ \cite{lattice,Hinchliffe} from lattice
computations.

We note that the BjPSR predictions deviate
from the world average in case ${\rm I}^{\prime}$
(\ref{BjdatIpr}), i.e., when we include in the
experimental data of case I the nuclear effects originating
from spin--one isosinglet 6--quark clusters in deuteron and helium
according to Ref.~\cite{SchmidtYang}, 
on top of the nuclear wavefunction effects
and NLO pQCD small--$x_{\rm Bj}$
extrapolation effects: 
${\alpha}_s^{{\overline {\rm MS}}}(M_Z^2) \approx
0.103^{+0.014}_{-0.027}$ (NNLO TPS);
$0.101^{+0.013}_{-0.025}$ (dPA, ECH, TPS--PMS, our approximant).
The combination of (older) case II results and
the mentioned 6--quark cluster nuclear effects 
(case ${\rm II}^{\prime}$) increases the value of the 
BjPSR integral so much that the predicted values of
${\alpha}_s^{\overline {\rm MS}}(M_Z^2)$
are unacceptably low: the central values would be
$0.094$--$0.095$ for all approximants;
the maximal allowed values would be about $0.113$
by the methods of Table \ref{tabl4} and
$0.114$ by the dPA.

The authors of Ref.~\cite{ABFR} obtained, among other things,
the BjPSR--predicted values
${\alpha}_s^{\overline {\rm MS}}(M_Z^2)\!=\!0.118^{+0.010}_{-0.026}$,
apparently using the simple NNLO TPS sum (\ref{TPSBj}) directly
in their analysis. They used the BjPSR--integral values
(\ref{BjdatII}), i.e.~here case II, which were extracted by them
from low--$Q^2_{\rm ph}$ SLAC experiments carried out before 1997. 
They used the 
value of $|g_A|\!=\!1.257$ known at the time, in contrast to
the value of (\ref{gA}). Their RGE evolution from
$Q^2_{\rm ph}\!=\!3{\rm GeV}^2$ to $M^2_Z$ was apparently
carried out at the three--loop level, since the
fourth--loop $\beta$--coefficient $c_3^{{\overline {\rm MS}}}(n_f)$
\cite{RVL} and the corresponding three--loop
flavor--threshold matching \cite{CKS} were not known at the time.
These two effects largely neutralize each other and their
result is then close to the NNLO TPS result for case II 
(Table \ref{tabl2}): 
${\alpha}_s^{\overline {\rm MS}}(M_Z^2)\!=\!0.118^{+0.010}_{-0.023}$.

The authors of Ref.~\cite{EGKS} obtained the BjPSR--predicted values
${\alpha}_s^{\overline {\rm MS}}(M_Z^2)\!=\!0.116^{+0.003}_{-0.005}\pm
0.003$. They used a dPA method of resummation $[2/2]_S$ mentioned
towards the end of Section IV. However, they took the
BjPSR--integral values (\ref{Regge}) where the naive
Regge small--$x_{\rm Bj}$ extrapolation was used, and apparently the
value $|g_A|\!=\!1.257$ known at the time. Further, they included
the effects of the higher--twist term (\ref{ht}) on top of
their dPA resummation. The additional uncertainty $\pm\!0.003$
can be called the method uncertainty. It was estimated by them
by additionally using the results of the nondiagonal
PA resummations $[1/2]_S$ and $[2/1]_S$,
the RScl--dependence of their dPA results, and the uncertainty
of the higher--twist term.

When we reexpand the approximants in powers of the original $a_0$
(at RScl $Q_0^2\!=\!Q^2_{\rm ph}$, in ${\overline {\rm MS}}$
RSch, $n_f\!=\!3$), we obtain predictions for coefficient $r_3$
at $a_0^4$ of expansion (\ref{TPSBj}) -- cf.~Eq.~(\ref{r3pr})
and the discussion following it. Our 
approximant, with $c_3\!=\!15.5$, predicts 
$r_3\!=\!125.8\!-\!c_3^{{\overline {\rm MS}}}/
2\!+\!c_3 \!\approx\!130.8$.
The ECH approximant, with $c_3\!=\!20.$,  
predicts $r_3\!=\!129.9\!+\! 
(-\!c_3^{{\overline {\rm MS}}}\!+\!c_3)/2\!\approx\!129.4$.
The two predictions are
close to each other, suggesting $r_3\!=\!130.\pm 1$. 
This agrees well with the prediction 
of Ref.~\cite{KS} 
$r_3\!\approx\!129.9$ ($\approx\!130.$)
which was obtained from the ECH under the assumption
$(-\!c_3^{{\overline {\rm MS}}}\!+\!c_3)\!\approx\!0$
(note that $c_3^{{\overline {\rm MS}}}\!\approx\!21.0$
\cite{RVL} was not even known at the time Ref.~\cite{KS} 
was written).

The predictions for $r_3$, as well as the values
of $Q_1^2$, $Q_2^2$, $c_2^{(1)}$, $c_2^{(2)}$
(\ref{ymres})--(\ref{c2res}) and of $c_3$ 
(Tables \ref{tabl1}, \ref{tabl3}), are for $n_f\!=\!3$
and are, of course,
independent of the specific values
for the BjPSR integral (\ref{BjdatI}), (\ref{BjdatII}) 
[$\Leftrightarrow$ (\ref{SI})--(\ref{SII})]
that we subsequently used to obtain values for
${\alpha}_s^{{\overline {\rm MS}}}(M_Z^2)$.

\section{Summary and outlook}

We presented an extension of our previous
method of resummation \cite{GC1,CK,GC2} for
truncated perturbation series (TPS) of massless QCD 
observables given at the next--to--next--to--leading order (NNLO).
While the previous method, partly related to the
method of the diagonal Pad\'e approximants (dPA's),
completely eliminated the unphysical dependence of the 
sum on the renormalization scale (RScl), the extension
presented here eliminates in addition the unphysical
dependence on the renormalization scheme (RSch).
The dependence on the leading RSch--parameter
$c^{(0)}_2\!\equiv\!{\beta}^{(0)}_2/{\beta}_0$ is eliminated
by a variant of the method of the principle of
minimal sensitivity (PMS). The dependence on the
next--to--leading RSch--parameter
$c^{(0)}_3\!\equiv\!{\beta}^{(0)}_3/{\beta}_0$ is eliminated
by fixing the $c_3$--value in the approximant
so that the correct value of
the location of the leading infrared renormalon (${\rm IR}_1$)
pole is obtained (by PA's of an RScl-- and RSch--invariant 
Borel transform). Hence, in the approximant we use 
$\beta$--functions which go beyond the highest calculated
order in the observable (NNLO) -- in order
to incorporate an important piece of nonperturbative information
(${\rm IR}_1$ pole location)
which is not contained in the available NNLO TPS anyway.
The results are apparently further
improved when we resum
those $\beta$--functions which are relevant for the
calculation of the approximant (RSch1 and RSch2 $\beta$--functions, 
for $a_1$ and $a_2$) and of 
${\alpha}_s^{{\overline {\rm MS}}}(Q_{\rm ph}^2)$
(${\overline {\rm MS}}$ RSch), by judiciously choosing
certain PA--forms for those $\beta$--functions.

We applied this method to the Bjorken polarized sum rule (BjPSR) 
at low values of the momentum transfer of the virtual photon 
$Q^2_{\rm ph}\!=\!5$ or $3 \ {\rm GeV}^2$.
The $c_3$--fixing by the ${\rm IR}_1$ pole location
is well motivated in this case, because the
contributions of the leading ultraviolet renormalon
(${\rm UV}_1$) appear to be sufficiently suppressed
in comparison to those of the ${\rm IR}_1$.
We compared predictions of our resummation with the
values for the BjPSR integral (\ref{BjdatI})
and (\ref{BjdatII}) extracted from experiments, and
obtained ${\alpha}_s^{\overline {\rm MS}}(M_Z^2)\!=\!
0.1187^{+0.0028}_{-0.0054}$ (new case I) and
$0.1127^{+0.0041}_{-0.0189}$ (older case II), respectively. 
Here, the central values 0.1187 and 0.1127 correspond
to the central values in (\ref{BjdatI}) and (\ref{BjdatII}),
respectively. For more discussion on the issue of the
experimental values (\ref{BjdatI}) and (\ref{BjdatII})
(cases I, II) we refer to Sections IV and VI.
It is gratifying that the newest available experimental values 
(\ref{BjdatI}) lead to predictions for
${\alpha}^{{\overline {\rm MS}}}_s$ which agree well
with the present world average.
The results of Grunberg's method
of the effective charge (ECH) and
of Stevenson's TPS--PMS method give very similar results
(cf.~Table \ref{tabl4})
if the $c_3$--parameter in these methods
is fixed by the same afore--mentioned requirement as
in our approximant and PA--forms of the pertaining
$\beta$--functions are chosen analogously. 
The combined result of Table IV, in case I, i.e.
with the newest data of Ref.~\cite{2000E155}, is
\begin{equation}
{\alpha}^{{\overline {\rm MS}}}_s(M_Z^2) = 
0.119^{+0.003}_{-0.006} \ .
\label{endresult}
\end{equation}
The dPA methods of resummation of $S$ predict higher values 
(central values about $0.120$ in case I; $0.114$ in case II), 
the non-diagonal PA's even higher 
(central values about $0.122$ in case I; $0.115$ in case II), 
and the NNLO TPS itself the highest values
(central value about $0.125$ in case I; $0.118$ in case II).

We expect that our approximant $\sqrt{{\cal A}_{S^2}}$, 
as well as the ECH and TPS--PMS, produced reliable resummation
results for the considered observable, because -- via their
dependence on $c_3$ -- we can incorporate into them 
in the afore--mentioned way important nonperturbative 
information about the ${\rm IR}_1$ pole,
and simultaneously achieve full RSch--independence.
The $c_3$--dependence in $\sqrt{{\cal A}_{S^2}}$,
in the ECH and in the TPS-PMS, 
is very closely related with the sensitivity of these approximants 
to the details of the corresponding RGE evolution.
These details ($c_3$--terms) in
the RGE evolution are numerically more important in
the lower energy regions, i.e., when the relevant energies
for the observable are low. Thus, significant $c_3$--dependence
of these approximants signals the relevance of 
nonperturbative regimes for the observable 
[cf. Eqs. (\ref{dAdc3})--(\ref{dAdc3TPS})]. 
It then appears natural that the $c_3$--parameter in these
approximants, i.e. the only parameter left free, 
is made to parametrize the location of
the (nonperturbative) ${\rm IR}_1$ pole.
The (d)PA's, in contrast, possess besides the $c_3$--dependence
also dependence on the leading RSch--parameter $c_2$,
and even on the RScl. Thus, the parameter $c_3$
in them is not in a special position, and there is
more ambiguity as to how to incorporate into the PA's
the information about the ${\rm IR}_1$ pole.

It appears that the leading higher--twist
term contribution to the BjPSR
($\sim$$1/Q^2_{\rm ph}$), or a part of it, is implicitly 
contained in $\sqrt{{\cal A}_{S^2}}$, as well as in the ECH and the
TPS--PMS, via the afore--mentioned $c_3$--fixing.
In this context, we point out that the so called
renormalon ambiguity arising from the ${\rm IR}_1$
of the BjPSR has the form
$\sim$$1/Q^2_{\rm ph}$, i.e., the form of the
leading higher--twist term. Even the coefficients of
this term, as estimated by the renormalon ambiguity
arguments, are of the same order of magnitude as
those predicted (estimated) from QCD sum rule.
One can say that the described approaches implicitly
give approximate--specific prescriptions for the elimination 
of the (leading IR) renormalon ambiguity.

Looking beyond the numerical analysis of the BjPSR,
we wish to stress that in cases of other
QCD observables that are (or eventually will be) known
to the NNLO, the analogous numerical analyses
may give different hierarchies of numerical results.
Actual resummation analyses should be performed also for such 
observables, in order to shed more light on the questions
about the relative importance of various kinds of
contributions.

The (d)PA methods, when applied directly to the (NNLO) TPS's,
are trying to include some nonperturbative
contributions through quasianalytic continuation
of the TPS from the perturbative (small--$a$)
to the nonperturbative (large--$a$) region.
In the course of this continuation,
the pole structure of the Borel transform
of the sum may be missed, but some other
nonperturbative (but less singular) features of the
sum itself may be reproduced well.
But our approximant $\sqrt{{\cal A}_{S^2}}$ would 
presumably do at least as good a job
as the dPA's in reproducing these latter
nonperturbative features. This is so because 
$\sqrt{{\cal A}_{S^2}}$ (\ref{Aans}) reduces to the 
dPA $[2/2]^{1/2}_{S^2}$ 
in the large--$\beta_0$ (one--loop RGE evolution)
approximation when thus the full RScl-- and RSch--invariance
requirements are abandoned -- cf. discussion following
Eqs.~(\ref{Aold})--(\ref{param2old}).
The ECH and the TPS--PMS
methods do not possess this strong ``$[2/2]^{1/2}$ PA--type'' 
mechanism of quasianalytic continuation,
since these two methods fix the RScl and
the RSch in the TPS itself without going beyond
the (NNLO) polynomial TPS form in $a$. The ECH, and
somewhat less explicitly the TPS--PMS, possess a weaker type 
of quasianalytic continuation, because the one--loop
RGE--evolved $a\!\equiv\!{\alpha}_s/\pi$ (from $a_0$) 
is a $[1/1]$ PA of $a_0$.

Stated differently, our (NNLO) approximants, from a
theoretical viewpoint, combine the favorable feature
of the (d)PA's (strong quasianalytic continuation
into the large--$a$ regime) with the favorable feature
of the TPS--form NNLO approximants ECH and TPS-PMS
(full RScl-- and $c_2$--independence). The residual
RSch--dependence ($c_3$--dependence) in the latter
approximants and in our approximant allows us
to incorporate into them, often in a well--motivated manner,
nonperturbative information on the location of the leading
IR renormalon pole, and to achieve in this way simultaneously
the full RSch--independence as well.

\acknowledgments

The work of G.C. was supported in part by the Korean Science
and Engineering Foundation.
We wish to acknowledge helpful discussion with
I.~Schmidt and J.-J.~Yang on the nuclear effects in the BjPSR.

\begin{appendix}

\section[]{Expansion of the general coupling 
\lowercase{$a$} in powers of \lowercase{$a_0$}}
\setcounter{equation}{0}
We outline here the derivation of the expansion of QCD
coupling $a\!\equiv\!a(\ln Q^2;c_2,c_3,\ldots)$ ($a=\alpha_s/\pi$)
in power series of 
$a_0\!\equiv\!a(\ln Q_0^2;c_2^{(0)},c_3^{(0)},\ldots)$.
The starting point is the Stevenson equation (cf. Ref.~\cite{PMS},
first entry, Appendix A)
which is obtained by integrating RGE (\ref{RGE})
\begin{equation}
\beta_0 \ln \left( \frac{Q^2}{{\tilde \Lambda}^2} \right) =
\frac{1}{a} + c_1 \ln \left( \frac{ c_1 a}{1\!+\!c_1 a} \right)
+ \int_0^a \!dx \left[ \frac{1}{x^2 ( 1\!+\!c_1 x)}
- \frac{1}{x^2 (1\!+\!c_1 x\!+\!c_2 x^2\!+\!c_3 x^3\!+\!\cdots )} \right] \ .
\label{Stevenson}
\end{equation}
It can be shown that ${\tilde \Lambda}$ here is
a universal scale ($\sim$$0.1$ GeV)
independent of the scale $Q$ and of the scheme parameters $c_j$
($j\!\geq\!2$). Writing the analogous equation for $a_0$,
and subtracting the two, we obtain
\begin{eqnarray}
\lefteqn{
\beta_0 \ln \left( \frac{Q^2}{Q_0^2} \right) = 
\frac{1}{a} + c_1 \ln \left( \frac{ c_1 a}{1\!+\!c_1 a} \right)
+ \int_0^a\!dx 
\frac{ (c_2\!+\!c_3 x\!+\!\cdots) }{( 1\!+\!c_1 x)
(1\!+\!c_1 x\!+\!c_2 x^2\!+\!c_3 x^3\!+\!\cdots )}
}
\nonumber\\
&& - \frac{1}{a_0} - c_1 \ln \left( \frac{ c_1 a_0}{1\!+\!c_1 a_0} \right)
- \int_0^{a_0}\!dx
\frac{ (c_2^{(0)}\!+\!c_3^{(0)} x\!+\!\cdots )}{( 1\!+\!c_1 x)
(1\!+\!c_1 x\!+\!c_2^{(0)} x^2\!+\!c_3^{(0)} x^3\!+\!\cdots )} \ .
\label{Stsubtr}
\end{eqnarray}
This equation determines $a$ as function of $a_0$. 
The solution $a$ in form of a power series of $a_0$ is
the Taylor series for function $a$ of multiple arguments
$\ln Q^2$ and $c_j$'s ($j\!\geq\!2$).
To obtain this power series, one way would be to find first
the derivatives 
$\partial a/\partial c_j$ [the derivative
$\partial a/\partial \ln Q^2$ is already given by RGE
(\ref{RGE})]. For this, we take the partial derivative
of both sides of the above equation with respect to $c_j$
($j\!\geq\!2$) and after some algebra we obtain 
\begin{eqnarray}
\frac{\partial a}{\partial c_j} &=&
a^2 ( 1 + c_1 a + c_2 a^2 + c_3 a^3 + \cdots )
\int_0^a \frac{dx x^{j-2}}{(1 + c_1 x + c_2 x^2 + c_3 x^3 + \cdots )^2} 
\ .
\label{intdercj}
\end{eqnarray}
Expanding the integrand in powers of $x$ and integrating out
each term, we obtain the partial derivatives as power series
\begin{eqnarray}
\frac{ \partial a}{\partial c_2} & = & 
a^3 \left( 1 + \frac{c_2}{3} a^2 + \cdots \right) \ ,
\label{derc2}
\\
\frac{ \partial a}{\partial c_3} & = & 
\frac{1}{2} a^4 \left( 1 - \frac{c_1}{3} a + \cdots \right) \ ,
\label{derc3}
\\
\frac{ \partial a}{\partial c_4} & = & \frac{1}{3} a^5 + \cdots \ .
\label{derc4} 
\end{eqnarray}
Repeated application of these equations, as well as
of RGE (\ref{RGE}) itself, leads us to the
following Taylor expansion of 
$a$ in powers of 
$a_0\!\equiv\!a(\ln Q_0^2;c_2^{(0)},c_3^{(0)},\ldots)$:
\begin{eqnarray}
a & = & a_0 + a_0^2 (-x) + a_0^3 ( x^2 - c_1 x + \delta c_2)
\nonumber\\
&& + a_0^4 ( - x^3 + \frac{5}{2} c_1 x^2 - c_2^{(0)} x
- 3 x \delta c_2 + \frac{1}{2} \delta c_3 ) 
\nonumber\\
&& + a_0^5 {\Big [} x^4 - \frac{13}{3} c_1 x^3 + 
( \frac{3}{2} c_1^2 + 3 c_2^{(0)} + 6 \delta c_2) x^2
\nonumber\\
&&+ ( - c_3^{(0)} - 3 c_1 \delta c_2 - 2 \delta c_3 ) x
+ ( \frac{1}{3} c_2^{(0)} \delta c_2 + 
\frac{5}{3} (\delta c_2)^2 - \frac{1}{6} c_1 \delta c_3 +
\frac{1}{3} \delta c_4 ) {\Big ]} + {\cal O}(a_0^6) \ ,
\label{aexpina0}
\end{eqnarray}
where we denoted
\begin{eqnarray}
a & \equiv & a(\ln Q^2;c_2,c_3,\ldots) \ , 
\quad
a_0 \equiv  a_0(\ln Q_0^2;c_2^{(0)},c_3^{(0)},\ldots) \ , 
\label{aa0}
\\
x &\equiv& \beta_0 \ln \frac{Q^2}{Q_0^2}, \quad
\delta c_k \equiv c_k - c_k^{(0)} \ .
\label{xdcj}
\end{eqnarray}

\section[]{Explicit $PMS$ conditions}
\setcounter{equation}{0}

Here we will write explicitly the PMS--like
conditions (\ref{PMSeqs}) in its lowest
order ($\sim$${\bar a_0}^5$). To do this, we calculate
explicitly the derivatives (\ref{PMSeqs})
and then expand them in powers of
${\bar a}_0\!=\!a(\ln Q_0^2; c_2=c_2^{(s)}; c_3; \ldots)$
to their lowest nontrivial order.\footnote{
In fact, $a$ with any RScl and any RSch--parameters
would do the job and give the same coefficient
at the leading nontrivial order $a^5$.}
We assume relation (\ref{dc30}), i.e.,
$\delta c_3\!=\!0$, and in addition
$\delta c_4$($\equiv\!c_4^{(1)}\!-\!c_4^{(2)}$)$=\!0$.
Further, we use relations (\ref{eq1})--(\ref{eq1b})
and notations (\ref{notf1})--(\ref{z0}).
The results, obtained with help of Mathematica,
are the following:
\begin{eqnarray}
\lefteqn{
\frac{ \partial {\cal A}_{\tilde S}^{[2/2]} }
{\partial c_2^{(s)} } {\Bigg |}_{\delta c_2} \equiv
- {\bar a}_0^5 {\bigg \{}
 27 (\delta c_2)^3 - 157 c_1 (\delta c_2)^2 y_{-}
- 8 \delta c_2 y_{-}^2 \left[ - 27 c_1^2\!+\!12 c_2^{(s)}
\!+\!34 y_{-}^2\!-\!8 z_0^2(c_2^{(s)}) \right]
}
\nonumber\\
&& + 48 c_1 y_{-}^3  
\left[ 13 y_{-}^2\!-\!3 z_0^2(c_2^{(s)}) \right]
{\bigg \}} {\bigg \{} 6 y_{-}^2 \left[
5 c_1 \delta c_2\!+\!16 y_{-}^3\!-\!8 z_0^2(c_2^{(s)}) y_{-} )
\right] {\bigg \}}^{-1} +\!{\cal O}({\bar a}_0^6) = 0 ,
\label{PMSeq1}
\\
\lefteqn{
\frac{ \partial {\cal A}_{\tilde S}^{[2/2]} }
{\partial (\delta c_2) } {\Bigg |}_{c_2^{(s)}} \equiv
- {\bar a}_0^5 {\bigg \{}
27 (\delta c_2)^4 - 315 c_1 (\delta c_2)^3 y_{-}
+ 64 z_0^4(c_2^{(s)}) y_{-}^2 \left[ 7 c_1^2\!-\!2 c_2^{(s)}
\!+\!3 z_0^2(c_2^{(s)}) \right]
}
\nonumber\\
&& - 80 c_1 \delta c_2 y_{-} \left[ - 2 c_2^{(s)} y_{-}^2
\!-\!2 c_2^{(s)} z_0^2(c_2^{(s)})\!+\!12 z_0^2(c_2^{(s)}) y_{-}^2
\!+\!3 z_0^4(c_2^{(s)})\!+\!7 c_1^2 
\left( y_{-}^2 \!+\!z_0^2(c_2^{(s)}) \right)
\right]
\nonumber\\
&& + 12 (\delta c_2)^2 \left[ - 2 c_2^{(s)} y_{-}^2
\!-\!2 c_2^{(s)} z_0^2(c_2^{(s)})\!+\!15 z_0^2(c_2^{(s)}) y_{-}^2
\!+\!3 z_0^4(c_2^{(s)})\!+\!c_1^2 
\left( 82 y_{-}^2\!+\!7 z_0^2(c_2^{(s)}) \right)
\right] {\bigg \}}
\nonumber\\
&&\times  {\bigg \{} 12 y_{-}^4 \left[
5 c_1 \delta c_2\!+\!16 y_{-}^3\!-\!8 z_0^2(c_2^{(s)}) y_{-}
\right] {\bigg \}}^{-1} + {\cal O}({\bar a}_0^6) = 0 \ ,
\label{PMSeq2}
\end{eqnarray}
The actual PMS--type equations are now obtained by requiring that
the coefficients at $\sim$${\bar a}_0^5$ in
(\ref{PMSeq1})--(\ref{PMSeq2}) be zero.
When we have several possible solutions of the coupled system
(\ref{eq1}) and (\ref{PMSeq1})--(\ref{PMSeq2})
for the three unknowns $y_{-}$, $c_2^{(s)}$ and $\delta c_2$, 
we have to choose, in the PMS--spirit, 
among the resulting approximants that
one which has the smallest curvature. The curvature
can be calculated by first obtaining the
eigenvalues $CA_1$ and $CA_2$ of the curvature
matrix $C_A$:
\begin{equation}
C_A =
\left[
\begin{array}{c c}
\frac{ {\partial}^2 {\cal A}_{\tilde S} }
{\partial (c_2^{(1)})^2 } & 
\frac{ {\partial}^2 {\cal A}_{\tilde S} }
{\partial c_2^{(1)} \partial c_2^{(2)} } \\
\frac{ {\partial}^2 {\cal A}_{\tilde S} }
{\partial c_2^{(1)} \partial c_2^{(2)} } &
\frac{ {\partial}^2 {\cal A}_{\tilde S} }
{\partial (c_2^{(2)})^2 }
\end{array} 
\right] \ ,
\label{curvmtrx}
\end{equation}  
\begin{equation}
{ CA_1 \choose CA_2 } =
\frac{1}{4} \frac{ \partial^2 {\cal A}_{\tilde S} }
{\partial (c_2^{(s)})^2 } +
\frac{ \partial^2 {\cal A}_{\tilde S} }
{\partial (\delta c_2 )^2 }
\pm \Bigg \{
\left( \frac{ \partial^2 {\cal A}_{\tilde S} }
{\partial (\delta c_2) \partial c_2^{(s)} } \right)^2
+ \left[ \frac{1}{4} \frac{ \partial^2 {\cal A}_{\tilde S} }
{\partial (c_2^{(s)})^2 } -
\frac{ {\partial}^2 {\cal A}_{\tilde S} }
{\partial (\delta c_2 )^2 }
\right]^2  \Bigg \}^{1/2} \ .
\label{curv12}
\end{equation}
In the last expression, we traded $c_2^{(1)}$ and
$c_2^{(2)}$ for $c_2^{(s)}\!\equiv\!(c_2^{(1)}\!+\!c_2^{(2)})/2$
and $\delta c_2\!\equiv\!(c_2^{(1)}\!-\!c_2^{(2)})$.
The curvature ${\cal C}_{\cal A}$ of the solution 
${\cal A}_{S^2}^{[2/2]}$
can be defined in at least two obvious ways which 
are virtually equivalent
\begin{equation}
{\cal C}_{\cal A} = |CA_1| + |CA_2| \ ,
\qquad {\rm or:} \quad
{\cal C}_{\cal A} = \sqrt{ (CA_1)^2 + (CA_2)^2 } \ .
\label{curv}
\end{equation}

\section[]{$ECH$ and $TPS$--$PMS$ methods for $NNLO$ $TPS$}
\setcounter{equation}{0}

The effective charge method (ECH) \cite{ECH} of
resummation of the NNLO TPS $S_{[2]}$ (\ref{S2TPS}) can
be expressed by employment of the subtracted
version (\ref{Stsubtr}) of Stevenson equation
\begin{eqnarray}
\lefteqn{
- r_1 + \frac{1}{a_0} + 
c_1 \ln \left( \frac{ c_1 a_0}{1\!+\!c_1 a_0} \right)
+ \int_0^{a_0}\!dx
\frac{ (c_2^{(0)}\!+\!c_3^{(0)} x + \cdots)}{( 1\!+\!c_1 x)
(1\!+\!c_1 x\!+\!c_2^{(0)} x^2\!+\!c_3^{(0)} x^3 + \cdots)} 
}
\nonumber\\
&&
= \frac{1}{a_{\rm ECH}} + c_1 
\ln \left( \frac{ c_1 a_{\rm ECH}}{1\!+\!c_1 a_{\rm ECH}} \right)
+ \int_0^{a_{\rm ECH}}\!dx 
\frac{ ({\rho}_2\!+\!c_3 x + \cdots) }{( 1\!+\!c_1 x)
(1\!+\!c_1 x\!+\!{\rho}_2 x^2\!+\!c_3 x^3 + \cdots)} \ .
\label{eqECH}
\end{eqnarray}
The ECH resummation value is $S^{\rm ECH}\!=\!a_{\rm ECH}$.
In (\ref{eqECH}), superscript ``$(0)$'' denotes 
the original RSch of $S_{[2]}$
(for example ${\overline {\rm MS}}$
RSch with $n_f\!=\!3$ in the considered BjPSR case),
and $c_3$ denotes the NNLO ECH value of $c_3$
(in principle unknown at NNLO). Further,
$c_2^{\rm ECH}\!=\!\rho_2$, the latter RScl-- and RSch--invariant 
is defined in (\ref{rho2}). 
The coupling $a_0\!\equiv\!{\alpha}_s^{(0)}/\pi$ is
defined $a_0 \equiv a(\ln Q_0^2; c_2^{(0)}, c_3^{(0)}, \ldots)$
as in (\ref{a0def1}), $Q_0^2$ being the original RScl
in the TPS (chosen equal $3 \ {\rm GeV}^2$ in
the considered BjPSR case);
$r_1\!=\!- \beta_0 \ln(Q^2_{\rm ECH}/Q_0^2)$ is the
NLO TPS coefficient as staying in (\ref{S2TPS}) at the
original RScl $Q_0^2$.
In the above relation (\ref{eqECH}), 
we often ignore the terms $\propto\!c_k^{(0)}$ and $c_k$
($k \geq 4$) since they are not known, i.e. we often choose
the TPS form for the $\beta(x)$--functions.
For a given value of $a_0$, solving the above relation
numerically for $a_{\rm ECH}$ gives us the resummed
prediction for observable $S$.
It is dependent on $c_3$ which, at this stage, is not known.
More explicitly:
\begin{equation}
S^{\rm ECH}(c_3) = a_{\rm ECH}(c_3) = 
a(\ln Q^2_{\rm ECH}; {\rho}_2, c_3, \ldots) \ ,
\quad {\rm with:} \ Q^2_{\rm ECH} = Q_0^2 \exp(- r_1/{\beta}_0) \ .
\label{eqECH2}
\end{equation}
For the TPS--PMS method \cite{PMS} applied to the
NNLO TPS $S_{[2]}$, relation (\ref{eqECH}) still remains valid,
but with the replacements 
\begin{equation}
a_{\rm ECH}(c_3) \mapsto a_{\rm PMS}(c_3) \ ,
\quad
c_2^{\rm ECH}\!\equiv\!\rho_2 \ \mapsto \ 
c_2^{\rm PMS}\!\equiv\!\frac{3}{2} \rho_2 \ .
\label{ECHPMS}
\end{equation}
The resummed expression in the (NNLO) TPS PMS case is the
following TPS:
\begin{equation}
S^{\rm PMS}(c_3) = a_{\rm PMS} - \frac{1}{2} \rho_2 a^3_{\rm PMS} \ ,
\quad {\rm with:}
\ a_{\rm PMS}(c_3) = a(\ln Q^2_{\rm ECH}; (3/2){\rho}_2, c_3, \ldots) \ ,
\label{SPMS}
\end{equation}
which again depends on $c_3$.
Expression (\ref{SPMS}) is obtained by imposing PMS
conditions on the TPS 
$S_{[2]}( \ln Q^2; c_2, c_3, \ldots)\!=\!S^{\rm PMS}$:
$\partial S_{[2]}/ \partial \ln Q^2 \sim a^5 \sim
\partial S_{[2]}/\partial c_2$.
It is straightforward to verify that, if $\rho_2 > 0$
(as in the considered BjPSR case), $S^{\rm PMS}$
is bounded from above due to its specific TPS form:
$S^{\rm PMS} \leq (2/3)^{3/2} \rho_2^{-1/2}$,
which in the considered BjPSR case (\ref{TPSBj}) is $0.2326$
(because ${\rho}_2\!=\!5.476$).

\end{appendix}

\noindent

\noindent

\noindent

\end{document}